\documentclass[12pt]{article}
\newtheorem{theorem}{Theorem}

\newtheorem{example}[theorem]{Example}

\usepackage{amssymb}
\def\binom#1#2{{#1 \choose #2}}
\newcommand{\dnth}{d^{N_T}\theta\;}
\newcommand{\te}{\theta}
\newcommand{\cas}{{\mbox{\footnotesize$\cal S$}}}
\newcommand{\dth}{\pa_\theta}
\newcommand{\tQ }{{\tilde Q}}
\newcommand{\bp}{{\bar\psi}}   \newcommand{\BP}{{\overline{\Psi}}}
\newcommand{\bfi}{{\bar\phi}}   \newcommand{\BFI}{{\overline{\Phi}}}
\newcommand{\bc}{{\bar{c}}}   \newcommand{\BC}{{\overline{C}}}
   
\newcommand{\bb}{{\bar b}}

\newcommand{\hO}{{\hat\Omega}}
\newcommand{\hA}{{\hat A}}

\newcommand{\hd}{{\hat d}}
\newcommand{\hF}{{\hat F}}

\makeatletter \@addtoreset{equation}{section} \makeatother

\def\ftoday{{\sl {Le \number\day \space\ifcase\month
\or janvier\or f\'evrier\or mars\or avril\or mai \or juin\or juillet\or
ao\^ut\or septembre\or octobre \or novembre \or d\'ecembre\fi\space
\number\year}}}
\def\ptoday{{\sl {\number\day \space de\space \ifcase\month
\or janeiro\or fevereiro\or mar{\c c}o\or abril\or maio \or junho\or
julho\or agosto\or setembro\or outubro \or novembro \or dezembro\fi\space
de\space \number\year}}}
\def\gtoday{{\sl {Den \number\day. \ifcase\month
\or Januar\or Februar\or M\"arz\or April\or Mai \or Juni\or Juli\or
August\or September\or Oktober \or November \or Dezember\fi\space
\number\year}}}
\def\today{{\sl {\ifcase\month
\or January\or February\or March\or April\or May \or June\or July\or
August\or September\or October \or November \or December\fi
\space\number\day,\space
                                            \number\year}}}
\newcommand{\journal}[4]{{\em #1~}#2\,(#3)\,#4}

\newcommand{\jhep}{\journal {J. High Energy Phys.}}

\newcommand{\ijmp}{\journal {Int. J. Mod. Phys.}}

\newcommand{\np}{\journal {Nucl. Phys.}}
\newcommand{\npproc}{\journal {Nucl. Phys. B (Proc.Suppl.)}}

\renewcommand{\a}{\alpha}

\renewcommand{\d}{\delta}         \newcommand{\D}{\Delta}
\newcommand{\e}{\epsilon}

\newcommand{\la}{\lambda}        \newcommand{\LA}{\Lambda}
\newcommand{\m}{\mu}
\newcommand{\n}{\nu}
\newcommand{\om}{\omega}         \newcommand{\OM}{\Omega}
\newcommand{\p}{\psi}             
           \renewcommand{\S}{\Sigma}

\newcommand{\f}{{\phi}}           \newcommand{\F}{{\Phi}}
\newcommand{\vf}{{\varphi}}
\newcommand{\XI}{\XI}
\newcommand{\CC}{{\cal C}}

\newcommand{\es}{\\[3mm]}

\newcommand{\sla}{\raise.15ex\hbox{$/$}\kern -.57em}
\newcommand{\Sla}{\raise.15ex\hbox{$/$}\kern -.70em}

\def\Lp{\displaystyle{\biggl(}}
\def\Rp{\displaystyle{\biggr)}}

\newcommand{\lp}{\left(}\newcommand{\rp}{\right)}
\newcommand{\lc}{\left[}\newcommand{\rc}{\right]}

\newcommand{\complex}{{\kern .1em {\raise .47ex
\hbox {$\scriptscriptstyle |$}}
    \kern -.4em {\rm C}}}
\newcommand{\real}{{{\rm I} \kern -.19em {\rm R}}}
\newcommand{\rational}{{\kern .1em {\raise .47ex
\hbox{$\scripscriptstyle |$}}
    \kern -.35em {\rm Q}}}
\renewcommand{\natural}{{\vrule height 1.6ex width
.05em depth 0ex \kern -.35em {\rm N}}}

\newcommand{\tr}{{\rm {Tr} \,}}
\newcommand{\half}{\dfrac{1}{2}}
\newcommand{\pa}{\partial}
\newcommand{\pad}[2]{{\frac{\partial #1}{\partial #2}}}

\newcommand{\dpad}[2]{{\displaystyle{\frac{\partial #1}{\partial
#2}}}}
\newcommand{\dfud}[2]{{\displaystyle{\frac{\delta #1}{\delta #2}}}}
\newcommand{\dfrac}[2]{{\displaystyle{\frac{#1}{#2}}}}
\newcommand{\dsum}[2]{\displaystyle{\sum_{#1}^{#2}}}
\newcommand{\dint}{\displaystyle{\int}}

\newcommand{\twiddle}{\lower.9ex\rlap{$\kern -.1em\scriptstyle\sim$}}

\newcommand{\vev}[1]{\left\langle {#1}\right\rangle}
\newcommand{\Vev}[1]{\displaystyle{\bigl\langle} {#1}\displaystyle{\bigr\rangle}}

\newcommand{\equ}[1]{(\ref{#1})}
\newcommand{\eq}{\begin{equation}}
\newcommand{\eqn}[1]{\label{#1}\end{equation}}
\newcommand{\eea}{\end{eqnarray}}
\newcommand{\eqa}{\begin{eqnarray}}
\newcommand{\eqan}[1]{\label{#1}\end{eqnarray}}
\newcommand{\ba}{\begin{array}}
\newcommand{\ea}{\end{array}}
\newcommand{\eqac}{\begin{equation}\begin{array}{rcl}}
\newcommand{\eqacn}[1]{\end{array}\label{#1}\end{equation}}

\begin{document}

%{\hfill\large {\bf n-top-fix.tex}\ \ \ptoday}
{\hfill\parbox{45mm}{{ 
hep-th/0310184\\
UFES-DF-OP2003/4\\
CBPF-NF-036/03
}} \vspace{3mm}
%%%%%%%%%%%%%%%%%%%%%%%%%%%%

\begin{center}
%{\LARGE\bf Generalized Blau-Thompson Gauge-Fixing with Batalin-Vilkovisky 
%Method in Superspace}
{\LARGE\bf  Superspace Gauge Fixing of \es
Topological Yang-Mills Theories}
\end{center}
\vspace{3mm}

%%%%%%%%%%%%%%%%%%%%%%%%%%%%%%%%
\begin{center}{\large 
Clisthenis P. Constantinidis$^{a}$\footnote{Supported 
in part by the Conselho Nacional 
de Desenvolvimento Cient\'{\i}fico e  
Tecnol\'{o}gico CNPq -- Brazil.}, 
Olivier Piguet$^{a,1}$ and 
Wesley Spalenza$^{b,1}$ }
\vspace{1mm}

%%%%%%%%%%%%%%%%%%%%%%%%%%%%%%%%%%%
\noindent %$^{*}$ 
 $^{a}$Universidade Federal do Esp\'{i}rito Santo (UFES)\\
$^{b}$Centro Brasileiro de Pesquisas F\'{i}sicas (CBPF)
\end{center}

%%%%%%%%%%%%%%%%%%%%%%%%%%%%%%%%%%%%%%%%%%

\begin{abstract}
We revisit the construction of 
topological Yang-Mills theories of the Witten type 
with arbitrary space-time dimension and number of ``shift supersymmetry'' 
generators, using a superspace formalism. 
The super-$BF$ structure of these theories is exploited in order
to determine their actions uniquely, up to the ambiguities
due to the 
fixing of the Yang-Mills and $BF$ gauge invariance. UV finiteness 
to all orders of
perturbation theory is proved in a gauge of the Landau type.
\end{abstract}

%%%%%%%%%%%%%%%%%%%%%%%%%%%%%%
%\tableofcontents

%%%%%%%%%%%%%%%%%%%%%%%%%%%%%%%%%%%%%%
\section{Introduction}

%%%%%%%%%%%%%%%%%%%%%%%%%%%%%%%%%%%%%%%%%%%%%%%%%%%%
Observables in topological  theories possess a global character,
such as the knot invariants of Chern-Simons theory, the Wilson
loops, etc. The problem of finding all theses invariants is a problem of
equivariant cohomology, as proposed by Witten in 
1988~\cite{witten-donald} for
Yang-Mills topological theory in four-dimensional space-time.
Equivariant cohomology is the cohomology of a BRST-like operator -- the
``shift supersymmetry operator'', associated to a local shift transformation
of the connection field -- in a space of gauge invariant field
polynomials. A superspace 
formulation of Witten's model was proposed by Horne~\cite{horne} and 
 developed later on, in particular by Blau and 
Thompson~\cite{bbrt-pr,blau-thom-96}, who extended it to the cases 
of more than 
one supersymmetry generator and in different space-time dimensions.
 In various cases these topological theories are seen to 
arise from super-Yang-Mills theories through some twist of 
group representations~\cite{witten-donald,bbrt-pr,blau-thom-96}, 
possibly accompanied by dimensional reduction. The reader may 
see~\cite{geyer} for the systematic construction of topological 
theories from super-Yang-Mills ones using this technique.
Our proposal is to systematize  the superspace construction of 
actions in the most general setting involving an arbitrary number 
$N_T$ of 
topological supersymmetry generators in any space-time dimension $D$. 
 Our construction will be direct, not passing through the twist 
procedure. The question of the existence, in each case, of a corresponding
super-Yang-Mills theory will not be touched.

The theory
intends to describe gauge field configurations with null curvature --
or also selfdual curvature, in the 4-dimensional case. The 
null or selfdual  curvature condition is implemented through a Lagrange 
multiplier field $B$ which has the same hierarchy of zero-modes as the 
$B$ field of a BF type theory~\cite{lucchesi-bf}. The point of view usually 
adopted in the 
literature~\cite{witten-donald,horne,bbrt-pr,blau-thom-96} 
is that of considering 
the supersymmetry generator(s) as BRST operator(s) associated to the local
shift invariance, and fixing the latter 
with suitable Lagrange multiplier fields. 
In the present paper, in order to avoid certain ambiguities which may
arise in the usual scheme, we shall
consider the theory as a rigid supersymmetric theory with two gauge
invariances, namely the usual Yang-Mills gauge invariance and the 
gauge invariance of the $B$ field, like the one encountered in $BF$-theories.
Both invariances are supergauge invariances, their parameters being 
superspace functions. We shall see that this is enough to define the
theory in an unambiguous way, apart from the freedom in the choice of a
gauge fixing procedure. Moreover, in the case of a gauge fixing
of the  Landau type, we
shall show,  using supergraph 
techniques, that perturbative radiative corrections are completely
absent. The theory thus turns out to be obviously ultraviolet finite.

 A very important point is the systematic characterization of all
observables of a topological theory. This has been 
fully done in~\cite{boldo}
for the $N_T=1$ theories. 
Partial results exist in the literature. In particular, a set of
observables has been given in~\cite{marculescu} for the  case of $N_T=2$
in a 4-dimensional K\"ahler manifold.
In the present paper we shall show a
rather general set of observables, for any value of $N_T$ and 
any space-time dimension,
however without determining if it
represents the most general set.

The plan of the paper is the following. After reminding the principal
features of the original Witten-Donaldson's topological Yang-Mills 
theory~\cite{witten-donald} in section \ref{shiftsusy} and of superspace 
formalism for topological theories in section \ref{ext-susy}, we shall 
show the construction of the action as a super-$BF$ one, with the 
appropriate gauge fixing, in section \ref{actions}. Examples of observables 
are given in section \ref{observables}, and the ultraviolert problem is 
dealt with in section \ref{uv-finiteness}. A discussion of the results
is done in the concluding section. Some of our conventions and
notations are given in two appendices.

%%%%%%%%%%%%%%%%%%%%%%%%%%%%%%%%%%%%%%%%%%%%
\section{Shift Supersymmetry}\label{shiftsusy}

We are going to review here, for illustrative purpose, 
how ``shift supersymmetry'' may describe the
gauge fixing of gauge field configurations with null curvature, 
or alternatively with selfdual curvature. We shall concentrate on the 
original Donaldson-Witten model~\cite{witten-donald,19}, 
with one supersymmetry generator
in four dimensional space-time. 

%%%%%%%%%%%%%%%%%%%%%%%%%%%%%%%%%%%%%%%%
\subsection{Transformation rules and invariant actions}

We recall that this model implies, 
beyond the gauge
connection $a_\m$ associated to some gauge group $G$,
a fermion 1-form $\p_\m$ and a 0-form $\f$, with
``shift'' supersymmetry defined by the infinitesimal
transformations
\eq\ba{l} 
\tQ  a = \p\ ,\quad \tQ  \p = -D(a)\f \equiv -\lp
d\f+[a,\f] \rp\ ,\quad \tQ \f=0\ .
\ea\eqn{N=1susy}
All fields here and in the rest of the paper are valued in the Lie 
algebra of the gauge group -- assumed to be a compact Lie group. Details
on the notation are given in appendix \ref{a-1}.

The usual Yang-Mills gauge transformations read, written as BRST 
transformations with ghost $c$:
\eq
\cas a=-D(a) c\ ,\quad \cas \p = -[c,\p]\ ,\quad \cas \f = -[c,\f]\ ,
\quad \cas c = -c^2\ .
\eqn{brst}
Whereas the BRST operator $\cas$ is nilpotent, the fermionic
generator $\tQ $ is nilpotent modulo a $\f$-dependent gauge transformation:
\eq\ba{l} 
\tQ ^2 a = -D(a)\f\ ,\quad \tQ ^2\p = -[\p,\f]\ ,\quad \tQ ^2\f=0\ .
\ea\eqn{tQ-nilpotency}
This means that $\tQ $ is nilpotent when restricted to gauge invariant quantities.
 Following Witten, we may thus interpret the shift supersymmetry invariance
as a BRST-like invariance in the space of gauge invariant field
functionals. The 1-form
$\p$ represents the ghost of local shift invariance and $\f$ is its
ghost of ghost. Then the following counting of
degrees of freedom holds: counting 4 degrees of freedom for $a$, $-4$ for
the ghost $\p$ and 1 for the ghost of ghost $\f$, we arrive to a total
of 1 degree of freedom, which corresponds to the scalar mode of the
field $a$ -- which in turn is eliminated thanks to the usual Yang-Mills gauge
invariance. The final number zero of local degrees of freedom is of course
characteristic of a topological theory.

In view of the absence of local degrees of freedom, the
theory may be defined through an action which will be purely of a gauge 
fixing type. This fixing of the local shift supersymmetry
may be done introducing
Lagrange multipliers fields  $\;^{0}{b}_2$, $\;^{1}\la_1$, 
$\;^{0}\la_0$ and $\;^{-1}\eta_0$, 
together with the corresponding ``antighost'' 
fields $\;^{-1}\bb_2$, $\;^{0}\bp_1$, 
$\;^{-1}\bp_0$ and $\;^{-2}\bfi_0$, where 
the lower right index denotes the form degree and the upper left one
the degree of supersymmetry or ``SUSY-number''. The latter number
corresponds to a ghost number in the interpretation of shift
supersymmetry as a BRST transformation.
Each ``antighost'' transforms under $\tQ $ into its corresponding Lagrange 
multiplier: 
\eq\ba{lll}
\tQ  \;^{-1}\bb_2 = \;^{0}b_2\ ,\quad& 
\tQ  \;^{p-1}\bp_p = \;^{p}\la_p\quad (p=1,2)\ ,\quad& 
\tQ  \;^{-2}\bfi_0 = \;^{-1}\eta_0\  , \es
\tQ  \;^{0}b_2 = -[ \;^{-1}\bb_2,\f]\ ,\quad&
\tQ  \;^{p}\la_p = -[ \;^{p-1}\bp_p,\f]\ ,\quad&
\tQ  \;^{-1}\eta_0 = -[ \;^{-2}\bfi_0,\f]\ ,
\ea\eqn{tQ-antigh}
the transformation rules of the Lagrange multipliers assuring the 
nilpotency of $\tQ $ modulo a $\f$-dependent gauge transformation. 
If one intends to study the instanton gauge field configurations, i.e.
those with selfdual
curvature $F=P_+F$, where $P_+$ is defined by \equ{selfduality}, 
the associated ``antighost'' and Lagrange mutiplier
have to be chosen as anti-selfdual: $P_+\;^{-1}\bp_0=0$, $P_+\;^{0}\la_0=0$.

A gauge invariant and $\tQ $-
invariant action may be taken as (following~\cite{blau-thom-96})
\eq\ba{l}
\tQ \;\tr\dint\Lp \;^{-1}\bb_2\; F(a) + \;^{-2}\bfi_0 \;D(a)*\p 
 + \;^{0}\bp_1 \;D(a)*^{-1}\bb_2 + \;^{-1}\bp_0 \;D(a)*^{0}\bp_1 \Rp \es
= \tr\dint \Lp \;^{0}{b}_2 F(a) + \;^{-1}\eta_0 \;D(a)*\p 
+ \;^{1}\la_1 \;D(a)*^{-1}\bb_2 + \;^{0}\la_0 \;D(a)*^{0}\bp_1\es
\phantom{= \tr\dint \Lp}
 + \;^{-1}\bb_2 \;D(a)\p + \;^{-2}\bfi_0 \;D(a)*D(a)\f 
 + \;^{0}\bp_1 \;D(a)*^{0}{b}_2 + \;^{-1}\bp_0 \;D(a)*^{1}\!\la_1\es
\phantom{= \tr\dint \lp\right.}
+ \;^{-2}\bfi_0 [\p,*\p] 
 - \;^{0}\bp_1 [\p,*^{-1}\bb_2] - \;^{-1}\bp_0 [\p,*^{0}\bp_1] \Rp\ ,
\ea\eqn{action-wz-n1d4}
where $F(a) = da+a^2$ is the curvature of the
Yang-Mills connection $a$ and $*$ is the Hodge duality operator
(see appendix \ref{a-1}). One sees that the 
Lagrange multiplier $\;^{0}{b}_2$ implements
the zero-curvature condition $F(a)=0$  -- or the selfduality condition, as 
in the original Witten's paper. $\;^{-1}\eta_0$  is the Lagrange multiplier
fixing the zero-mode of $\p$. Morover, $\;^{1}\la_1$ fixes
the zero-mode of $\;^{-1}\bb_2$, $\;^{0}\la_0$ that of $\;^{0}\bp_1$.
Finally, $\;^{0}\bp_1$ fixes the zero-mode of $\;^{0}{b}_2$ and 
$\;^{-1}\bp_0$ that of $\;^{1}\la_1$. 

This action corresponds to a generalized ``Landau gauge'' fixing. However,
it is still possible to add one invariant term quadratic in the 
Lagrange multipliers without spoiling gauge invariance, supersymmetry and
SUSY-number conservation. It reads
\eq
\tQ \;\tr\dint \dfrac{\xi}{2}\;^{-2}\bp_2*\;^0{b}_2
=\tr\dint \dfrac{\xi}{2}\lp \;^0{b}_2 *\;^0{b}_2
+ \;^{-2}\bp_2 *[\;^{-2}\bp_2,\f] \rp\ .
\eqn{quad-term-n1d4}
In this case, which corresponds to a generalized Feynman gauge,
 the Lagrange mmultiplier $\;^0{b}_2$ becomes an auxiliary fields, whose 
elimination through its equation of motion
gives rise, for $\xi=1$, to the original action of Witten~\cite{witten-donald}, 
which describes selfdual configurations when choosing the 
Lagrange multiplier $\;^0{b}_2$ and its corresponding ``antighost'' as
anti-selfdual 2-forms.

%%%%%%%%%%%%%%%%%%%%%%%%%%%%%%%%%%%%%%%%
\subsection{Observables}\label{observ-de-Witten}

According to Witten, the algebra of observables of the theory is generated
by  sets of gauge invariant forms $w_p^{(n)}$ ($0\le p\le4$, $n$ integer) 
obeying "descent equations"
\eq
\tQ w_p^{(n)} + d w_{p-1}^{(n)}= 0\quad (4\ge p\ge1)\ ,\quad
\tQ w_0^{(n)} =0\ ,
\eqn{witten-observ}
and are uniquely fixed up to total derivatives by
\[
w_0^{(n)} = C^{(n)}(\f)\ ,%\tr \f^n\ .
\]
where $C^{(n)}(\f)$ is an invariant corresponding to a Casimr operator 
$\CC^{(n)}$ of the gauge group.
Each $p$-form $w^n_p$ being then integrated on some $p$-dimensional submanifold
$M_p$, represents an equivariant cohomology class and defines a basis
element of the algebra of observables.

%%%%%%%%%%%%%%%%%%%%%%%%%%%%%%%%%%%%%%%%%
\section{$N_T$-Extended Supersymmetry}\label{ext-susy}

Our purpose in this section is to review and develop a superspace formalism
describing topological theories such as Wittens's theory described in 
section \ref{shiftsusy} and generalizations of 
it for more than one supersymmetric 
generators and for any space-time dimension, starting from the formalism
described in \cite{horne,blau-thom-96,boldo}.

%%%%%%%%%%%%%%%%%%%%%%%%%%%%%%%%%%%%%%%%%%%%
\subsection{$N_T$ Superspace formalism}

$N_T$ supersymmetry is generated by the fermionic charges 
$Q_I$, $I=1,\dots,N_T$ obeying the Abelian 
superalgebra \footnote{The bracket is here an anti-commutator.}
\eq
[Q_I,Q_J] = 0\ ,
\eqn{s-agebra|}
commuting with the space-time 
symmetry generators and the gauge group generators. The gauge group is some
compact Lie group.

A representation of supersymmetry is provided by superspace, a
supermanifold with $D$  bosonic and $N_T$
fermionic dimensions\footnote{Notations and conventions on
superspace are given in appendix \ref{a-2}.}. 
The respective coordinates are denoted by 
$(x^{\mu }$, $\mu=0,\dots,D-1)$, and  ($\theta ^{I}$,  $I=1,\dots,N_T)$. 
A superfield is by definition a superspace function  $F(x,\theta )$
which transforms as
\begin{equation}
Q_{I}F(x,\theta ) = \partial_{I}F(x,\theta )
\equiv  \frac{\partial }{\partial \theta ^{I}}F(x,\theta )
\eqn{s-field-transf}
under an infinitesimal supersymmetry transformation.

An expansion in the coordinates $\theta ^{I}$ of a generic superfield
reads
\begin{equation}
F(x,\theta )=f(x)+\dsum{n=1}{N}\frac{1}{n!}\theta
^{I_{1}}...\theta ^{I_{n}}f_{I_{1}...I_{n}}(x)  
\eqn{s-field-exp}
where the space-time fields $f_{I_{1}...I_{n}}(x)$ are completely
antisymmetric in the indices $I_{1}...I_{n}$. We recall that all fields 
(and superfields) are Lie algebra valued.
We shall also deal with superforms. A $p$-superform may be written as
\eq
\hO_p = \sum_{k=0}^{p}  %\dsum{k=0}{p} 
\OM_{p-k;\,I_1...I_k} d\theta^{I_1}\cdots d\theta^{I_k}\ ,
\eqn{s-form}
where the coefficients $\OM_{k-1;\,I_1...I_k}$ are (Lie algebra valued)
superfields which are space-time forms of degree $(p-k)$. They are
completely symmetric in their indices since, the
coordinates $\theta $ being anti-commutative, the
differentials $d\theta^I $ are commutative. 
The superspace exterior derivative is defined as
\begin{equation}
\hd=d+d\theta ^{I}\partial _{I}\ ,\quad d=dx^{\mu }\partial _{\mu }\ ,
\end{equation}
and is nilpotent: $\hd\;^{2}=\;0$. 

The basic superfield of the theory is the superconnection $\hA$, 
a $1$-superform:
\begin{equation}
\hA=A+E_{I}d\theta ^{I}  \ ,
\label{superconexao}\end{equation}
with $A=A_{\mu }(x,\theta )dx^{\mu }$ a $1$-form superfield and 
$E_{I}=E_{I}(x,\theta )$ a $0$-form superfield. The
superghost $C(x,\theta)$ is a $0$-superform. 
 We expand the components of the superconnection (\ref{superconexao}) as
\begin{equation}
A=a(x)+ \dsum{n=1}{N}\frac{1}{n!}\theta
^{I_{1}}...\theta ^{I_{n}}a_{I_{1}...I_{n}}(x)\ ,
\label{expansion-A}\end{equation}
where the 1-form $a$ is the gauge connection, and the 1-forms 
$a_{I_{1}...I_{n}}$ its supersymmetric partners.
The expansions of $E_{I}$ and of the ghost superfield $C$ read
\eq\ba{l}
E_{I}=e_{I}(x)+\dsum{n=1}{N} \frac{1}{n!}\theta
^{I_{1}}...\theta ^{I_{n}}e_{I,I_{1}...I_{n}}(x)\ ,\es
C=c(x)+\dsum{n=1}{N}\frac{1}{n!}\theta
^{I_{1}}...\theta ^{I_{n}}c_{I_{1}...I_{n}}(x)\ .
\ea\eqn{expansions-E-C}
The infinitesimal supergauge transformations of the superconnection 
are expressed as the nilpotent BRST transformations
\begin{equation}
\cas \hA=-\hd C-[C,\hA]\ ,\quad \cas C=-C^{2}\ ,\quad \cas ^2=0\ .
\label{BRST-Transform}
\end{equation}
In terms of component superfields we have
\begin{equation}
\cas A=-dC-[C,A]\ ,\quad \cas E_{I}=-\partial _{I}C-[C,E_{I}]\ ,
\quad \cas C=-C^{2}\ .
\label{BRST-comp}\end{equation}
The supercurvature 
\eq
\hF=\hd\hA + \hA^2 = F(A) + \Psi_I \;d\theta^I 
+ \F_{IJ}\;d\theta^Id\theta^J
\eqn{superFStrenght}
transforms covariantly:
\[
\cas \hF = -[C,\hF]\ ,
\]
as well as its components
\eq
F(A) = dA+A^{2}\ ,\quad \Psi_I = \partial_{I}A + D(A)E_{I}\ ,
\quad \F_{IJ} = \half\lp \pa_{I}E_{J}+\pa_{J}E_{I} + [E_{I},E_{J}] \rp\ ,
\eqn{superComp}
where the covariant derivative is defined by
$D(A)(\cdot )=d(\cdot )+[A,(\cdot )]$

%%%%%%%%%%%%%%%%%%%%%%%%%%%%%%

For further use and comparisons with the literature, let us give 
the explicit examples of $N_T=1,\,2$.

%%%%%%%%%%%%%%%%%%%%%%%%%%%%%%%%%%%%%
%\subsubsection*{Case $N_T=1$:}

\begin{example} {\bf -- Case $N_T=1$}

The superconnection \equ{superconexao} and the expansions 
(\ref{expansion-A} - \ref{expansions-E-C}) read
\eq
A(x,\theta ) = a(x)+\theta \psi (x) \ ,\quad
E(x,\theta ) = \chi (x)+\theta \phi (x) \ ,\quad
C(x,\theta ) = c(x)+\theta c'(x)\ .
\eqn{exp-N=1}
The BRST\ transformations of the component fields are
\eq\ba{llll}
\cas a = -D(a)c \ ,\ 
&\cas \psi  =-[c,\psi ]-D(a)c'\ ,\ 
&\cas \phi  =-[c,\phi ]-[\chi ,c']\ ,\ 
\cas \chi  =-[c,\chi ]-c'\ ,\es
\cas c = -c^{2} \ ,\ 
&\cas c' =-[c,c'] &&
\ea\eqn{trsfoBRSTN1}
As for the supersymmetry transformations defined by 
\equ{s-field-transf}, we have:
\eq
Qa = \psi\ ,\ Q\psi  = 0 \ ,\qquad 
Q\chi  = \phi\ ,\ Q\phi  = 0  \ ,\qquad
Qc = c'\ ,\ Qc' = 0  \ .
\eqn{transfQ}
The supercurvature components \equ{superComp} read 
\eq\ba{l}
F(A) = F(a) -\theta D(a)\p\ ,\es
\Psi = \p +D(a)\chi - \theta\lp D(a)\f - [\p,\chi] \rp\ ,\es
\Phi = \f +\chi^2 +\theta [\f,\chi]\ .
\ea\eqn{supercampo3}
\end{example}

%%%%%%%%%%%%%%%%%%%%%%%%%%%%%
%%%%%%%%%%%%%%%%%%%%%%%%%%%%%%%%%%%%%
%\subsubsection*{Case $N_T=2$:}
\begin{example} {\bf -- Case $N_T=2$}

The superconnection \equ{superconexao} and the expansions 
(\ref{expansion-A} - \ref{expansions-E-C}) now read (with $I=1,2$)
\eq\ba{l}
A(x,\theta ) = a(x)+\theta^I \psi_I (x) 
+\frac{1}{2}\theta ^{2}\alpha\ ,\quad
E_I(x,\theta ) = \chi_I (x)+\theta^I \phi_{IJ} (x)
+\frac{1}{2}\theta ^{2}\eta _{I} \ ,\es
C(x,\theta ) = c(x)+\theta^I c_I(x) +\frac{1}{2}\theta ^{2}c_{F} \ .
\ea\eqn{exp-N=2}
The BRST transformations of the component fields are
\begin{equation}\begin{array}{l}
\cas a= -D(a)c\ , \quad
\cas \psi _{I}=-[ c,\psi _{I}] -D(a)c_{I}, \quad
\cas \alpha = -[c,\alpha ]-D(a)c_{F}+\epsilon ^{IJ}[ c_{I},\psi _{J}]\ ,\es
\cas \chi _{I}=-[c,\chi _{I}]-c_{I}\ , \quad
\cas \phi _{IJ}=-[c,\phi _{IJ}]-\epsilon _{IJ}c_{F}+[\chi _{I},c_{J}]\ ,
\es
\cas \eta _{I}=-[c,\eta _{I}]-[c_{F},\chi _{I}]+\epsilon ^{JK}[c_{J},\phi
_{IK}]\ , \es
\cas c=-c^{2}\ , \quad
\cas c_{I}=-[c,c_{I}]\ , \quad
\cas c_{F}=-[c,c_{F}]+\frac{1}{2}\epsilon ^{IJ}[ c_{I},c_{J}]\ .
\end{array}\label{trsfoBRSTN2}\end{equation} 
The supersymmetry transformations read
\eq\ba{lll}
Q_{I}a=\psi _{I}\ ,&\quad Q_{I}\psi _{J}=-\epsilon _{IJ}\alpha \ ,&\quad
Q_{I}\alpha =0\ ,\es
Q_{I}\chi _{J}=\phi _{JI}\ , &\quad Q_{I}\phi _{Jk}=-\epsilon _{IK}\eta _{J}
  \ ,&\quad Q_{I}\eta _{J}=0\ ,\es
Q_{I}c=c_{I}\ , &\quad Q_{I}c_{I}=-\epsilon _{IJ}c_{F}\ ,&\quad
Q_{I}c_{F}=0  \ .
\ea\eqn{transfQ-N=2}
The supercurvature components \equ{superComp} read now
\eq\ba{l}
F(A) = F(a) -\theta ^{I}D(a)\psi _{I}+\frac{1}{2}\theta ^{2}(D(a)\alpha 
-\frac{1}{2}\epsilon ^{IJ}\left[ \psi _{I},\psi _{J}\right] ) \ ,\es
\Psi_I = \p_I +D(a)\chi_I +\theta^{J}(\epsilon_{IJ}\alpha -
D(a)\phi _{IJ}+[\psi _{J},\chi _{I}])  
\es\phantom{F(A) =}
+\frac{1}{2}\theta ^{2}(D(a)\eta _{I}-\epsilon ^{KJ}[\psi
_{K},\phi _{IJ}]+[\alpha ,\chi _{I}])\ ,\es
\Phi_{IJ} = \half \lp \f_{IJ}+\f_{JI} +[\chi_I,\chi_J]  
+\theta ^{K}(\epsilon_{IK}\eta _{J} + \epsilon_{JK}\eta _{I}
+[\phi _{JK},\chi_{I}] + [\phi _{IK},\chi _{J}])\right. 
\es\phantom{F(A) =}
+\left. \frac{1}{2}\theta ^{2}( [\chi_{I},\eta _{J}] + [\eta _{I},\chi _{J}]
- \epsilon ^{KL}[\phi _{IK},\phi _{JL}])\rp\ .
\ea\eqn{supercampo3-N=2}
\end{example}

%%%%%%%%%%%%%%%%%%%%%%%%%%%%%%%%%%%%%%%%%%%
\subsubsection*{Counting the number of degrees of freedom:}

The numbers of degrees of freedom, i.e. the numbers of component fields --
remembering that a $p$-form has $D!/[p!(D-p!)]$ components -- 
are shown in Table \ref{tab-dof}.
\begin{table}[hbt]
\centering
\begin{tabular}{|l||c|c|c|c|c|c|}    %c|c|c|}
\hline
Fields:  & $a_{I_1\cdots I_n}(x)$ & $A(x,\theta)$ 
& $e_{I,I_1\cdots I_n}(x)$ &$E_I(x,\theta)$
         & $c_{I_1\cdots I_n}(x)$ & $C(x,\theta)$ \\ \hline\hline 
$\ba{c}\mbox{Numbers}\\ \mbox{of fields:}\ea$
&$D\binom{N_T}{n}$  &$D\; 2^{N_T}$ &$N_T\binom{N_T}{n}$ 
&$N_T\;2^{N_T}$ &$\binom{N_T}{n}$ &$2^{N_T}$\\   \hline 
\end{tabular}
\caption[t1]{Numbers of component fields. $D$ = space-time dimension, 
$N_T$ = number of supersymmetry generators.}
\label{tab-dof}
\end{table}
If we were considering the present theory as a usual supersymmetric gauge theory, 
with (super)gauge
invariance defined by the BRST transformations \equ{BRST-comp}, the 
number of physical degrees of freedom would be given by the total number
of components of the forms $A$ and $E_I$ minus the number of components
of the superghost $C$. However, considering it as a topological theory we
have to treat supersymmetry as a local invariance, too, all fields excepted 
the Yang-Mills connection $a$ being
ghosts or ghosts of ghosts, as in the example shown in section
\ref{shiftsusy}. The SUSY-number $s$ is thus a ghost number as well as 
the usual ghost number\footnote{$s$ and $g$ are defined by attributing 
$s=g=0$ to the gauge connection $a(x)$, $s=1,\,g=0$ for 
the supersymmetry genrators $Q_I$ -- hence $s=-1$ to $\theta^I$ --
and $s=0,\,g=1$ for the BRST generator $\cas$.}
 $g$. Thus the effective ghost number is equal to $s+g$ and, in the
counting of the physical degrees of freedom, we must therefore assign a 
sign $(-)^{s+g}$ to the number of degrees of freedom of a field, as shown 
in Table \ref{tab-ph-dof}.
\begin{table}[hbt]
\centering
\begin{tabular}{|c||c|c|c|c|c|c|}    %c|c|c|}
\hline
Fields:  & $a_{I_1\cdots I_n}$ & $A$ 
& $e_{I,I_1\cdots I_n}$ &$E_I$
         & $c_{I_1\cdots I_n}$ & $C$ \\ \hline\hline
SUSY \#: &$n$ &$0$ &$n+1$ &$1$ &$n$ &$0$ \\ \hline
Ghost \#: &$0$ &$0$ &$0$ &$0$ &$1$ &$1$ \\ \hline
% &&&&&&\\
$\ba{c}\mbox{Degrees of} \\ \mbox{freedom:}\ea$ 
&$(-1)^n D\binom{N_T}{n}$  &$0$ 
&$(-1)^{n+1}N_T\binom{N_T}{n}$ 
&$0$ &$(-1)^{n+1}\binom{N_T}{n}$ &$0$\\  %&&&&&&\\ 
\hline 
\end{tabular}
\caption[t1]{Numbers of physical degrees of freedom. $D$ = space-time dimension, 
$N_T$ = number of supersymmetry generators.}
\label{tab-ph-dof}
\end{table}
One sees that there is a complete cancellation of the local degrees of
freedom, as it should in a topological theory.

%%%%%%%%%%%%%%%%%%%%%%%%%%%%%%%%%%%%%%%%%%%%%%%%
\subsection{Wess-Zumino gauge}

The contact with the formalism described in section \ref{shiftsusy} is
made by choosing a special gauge fixing~\cite{blau-thom-96} 
of the Wess-Zumino type~\cite{1001}.
The BRST transformations  of the component fields can be calculated from 
the superfield expressions \equ{BRST-comp}. They are explicitly given,
 for $N_T$ = 1 and 2, by (\ref{trsfoBRSTN1},\ref{trsfoBRSTN2}).
We shall only write explicitly the linear part -- or Abelian 
approximation -- of the
transformations in the general case, which will be sufficient for 
our argument:
\eq\ba{ll}
\cas a =-dc+\cdots\ ,&\quad
\cas a_{I_{1}...I_{N}}=-dc_{I_{1}...I_{N}}+\cdots\ ,\es
\cas e_{I} =-c_{I}+\cdots\ ,&\quad
\cas e_{I,I_{1}...I_{N}}=-c_{II_{1}...I_{N}}+\cdots\ ,\es
\cas c =\cdots\ ,&\quad
\cas c_{I_{1}...I_{N}}=\cdots\ ,
\ea\eqn{brst-n-lin}
where the dots represent nonlinear terms.
These transformaions indicate that $e_I(x)$ and the completely
antisymmetric part of the fields $e_{I,I_{1}...I_{n}}(x)$ are pure
gauge degrees of freedom. A possible gauge fixing is therefore of
setting these fields to zero. This defines the Wess-Zumino (WZ) gauge:
\begin{equation}
e_{I}=0\ ,\quad e_{[I,I_{1}...I_{n}]}=0\ (1\leq n\leq N_T)\ .  
\label{condi-NWZ}\end{equation}
This fixes the gauge degreees of freedom corresponding to 
the ghosts 
$c_{I_{1}...I_{n}}$ ($1\leq n\leq N_T$). The remaining gauge degree of freedom 
parametrized by the ghost $c$, which is of the usual Yang-Mills 
type, can be fixed in a usual way.

The WZ gauge condition (\ref{condi-NWZ}) is not stable under supersymmetry
transformations, but one can redefine the generators $Q_I$ into new
generators $\tQ _I$, compatible with the WZ condition, resulting from a 
combination of $Q_I$ and of a field dependent supergauge transformation. 
Thus, let us combine an infinitesimal 
supersymmetry transformation of constant commuting parameters $\e^I$ with a
supergauge transformation $\delta _{\Lambda }$
of anticommuting parameters (fermionic superfield) $\LA(x,\theta)$:
\begin{equation}
\tQ = \e^IQ_I   +\delta _{\Lambda } \equiv \e^I\tQ_I \ .
\label{condigaugeN}\end{equation}
 $\delta_{\Lambda }$ is in fact a BRST transformation (\ref{trsfoBRSTN1}), with
$C$ substituted by $\Lambda $. This will define the modified supersymmetry
generator $\tQ_I$, provided we choose $\LA$ in such a way to preserve the
WZ gauge condition (\ref{condigaugeN}).
It is convenient to rewrite the WZ
condition in a superspace way:
\begin{equation}
\theta ^{I}E_{I}(x,\theta )=0 \ , 
\label{condicaoW-Z}
\end{equation}
where $E_{I}$ is the $d\theta$-part of the superconnection
\equ{superconexao}.
We shall denote by $\tilde{E}_{I}$ the solution of this condition, and by 
$\tilde{e}_{[I,I_{1}...I_{n}]}$ ($0\leq n\leq N_T$) the components of
its $\theta$-expansion -- which are therefore solutions of \equ{condi-NWZ}. The
latters are tensors with mixed symmetry.
Applying $\tQ $ to \equ{condicaoW-Z} we find, after some integration by part
in $\theta$:
\begin{equation}
\tQ (\theta ^{I}E_{I})= -\theta^I\lp \e^J\pa_J E_I -\pa_I C - [C,E_I]\rp 
=-\epsilon ^{I}E_{I}+\theta ^{I}\partial
_{I}\Lambda +\partial _{J}(\epsilon ^{J}\theta ^{I}E_{I})+[\theta
^{I}E_{I},\Lambda ] \ ,
\end{equation}
which shows that the WZ condition (\ref{condicaoW-Z}) is stable
if, and only if, $\Lambda $ obeys the equation
\begin{equation}
\theta ^{I}\partial _{I}\Lambda =\epsilon ^{I}E_{I}\ .
\end{equation}
The solution reads
\begin{equation}
\Lambda =\epsilon ^{I}\sum_{n=1}^{N_T}\frac{1}{n!n}
\theta ^{I_{1}}\cdots\theta^{I_{n}}\tilde{e}_{I,I_{1}\dots I_{n}}(x)\ ,
\end{equation}
where the functions $\tilde{e}_{I,I_{1}...I_{n}}(x)$  are the coefficients 
of the superfield expansion of $\tilde{E}_{I}$, solution of 
(\ref{condicaoW-Z}).

One can now check that the superalgebra now closes up to field dependent
gauge transformations $\d_{\displaystyle{\tilde e_{IJ}}}$:
\[
[\tQ_I,\tQ_J] = -2 \d_{\displaystyle{\tilde e_{IJ}}}\ .
\]

%%%%%%%%%%%%%%%%%%%%%%%%%%%%%%%%%%%%%%%%%
\subsubsection*{Physical degrees of freedom in the WZ gauge:}

The numbers of component fields are now given in Table \ref{tab-dof-WZ}.
Remember that the only remaining ghost is $c(x)$, 
since the $c_{I_1\dots I_n}$
for $n\ge1$ correspond to the gauge degrees of 
freedom which have been fixed.
\begin{table}[hbt]
\centering
\begin{tabular}{|l||c|c|c|c|c|}
\hline
Fields:  & $a_{I_1\cdots I_n}(x)$ & $A(x,\theta)$ 
& $e_{I,I_1\cdots I_n}(x)$ &$E_I(x,\theta)$
         & $c(x)$  \\ \hline\hline 
$\ba{c}\mbox{Numbers}\\ \mbox{of fields:}\ea$
&$D\binom{N_T}{n}$  &$D\; 2^{N_T}$ 
&$\binom{N_T+1}{n+1} n$ 
&$(N_T-1)2^{N_T}+1$ &$1$ \\   \hline 
\end{tabular}
\caption[t1]{Numbers of component fields in the WZ gauge. 
$D$ = space-time dimension, 
$N_T$ = number of supersymmetry generators.}
\label{tab-dof-WZ}
\end{table}
In order to count the physical degrees of freedom we must again take 
into account the 
sign $(-)^{s+g}$ characterizing the ghost nature of each field, thus 
obtaining the results shown in Table \ref{tab-ph-dof-WZ}.
\begin{table}[hbt]
\centering
\begin{tabular}{|c||c|c|c|c|c|}
\hline
Fields:  & $a_{I_1\cdots I_n}(x)$ & $A(x,\theta)$ 
& $e_{I,I_1\cdots I_n}(x)$ &$E_I(x,\theta)$
         & $c(x)$ \\ \hline\hline
SUSY \#: &$n$ &$0$ &$n+1$ &$1$ &$0$  \\ \hline
Ghost \#: &$0$ &$0$ &$0$ &$0$ &$1$  \\ \hline
% &&&&&&\\
$\ba{c}\mbox{Degrees of} \\ \mbox{freedom:}\ea$ 
&$(-1)^n D\binom{N_T}{n}$  
&$0$ 
&$(-1)^{n+1}\binom{N_T+1}{n+1} n$ 
&$1$ &$-1$ \\  %&&&&&&\\ 
\hline 
\end{tabular}
\caption[t1]{Numbers of physical degrees of freedom. $D$ = space-time dimension, 
$N_T$ = number of supersymmetry generators.}
\label{tab-ph-dof-WZ}
\end{table}
There is again a complete cancellation of the local degrees of
freedom, as it should.

%%%%%%%%%%%%%%%%%%%%%%%%%%%%%%%%%%%%%%%%%%
Let us consider more explicitly the cases of $N_T$ = 0 and 1.

\begin{example} {\bf -- Case $N_T=1$}

The WZ gauge condition reads $\chi=0$, we have 
\[
\tilde E(x,\theta) = \theta\f(x)\ ,
\]
and the parameter $\LA$ of the compensating supergauge transformation is
given by
\[
\LA= \e\; \theta \f \ .
\]
In 4 dimensions we 
recover the Donaldson-Witten theory of section \ref{shiftsusy}.
In particular, the modified supersymmetry transformations are those given
by \equ{N=1susy}. It is moreover easy to check the nilpotency of $\tQ$
modulo a $\f$-dependent gauge tansformation $\d_\f$:
\[
\tQ^2 = \d_\f\ .
\]
\end{example}

%%%%%%%%%%%%%%%%%%%%%%%%%%%%%%%%%%%%
\begin{example} {\bf -- Case $N_T=2$}

In terms of the component fields defined by \equ{exp-N=2},
the WZ gauge condition reads
\[
\chi_I=0\ ,\quad \f_{IJ} - \f_{JI} = 0\ ,
\]
so that
\[
\tilde{E}_I = \theta^J \f_{(IJ)} + \half\theta^2 \eta_I\ ,\quad
\mbox{with}\quad \f_{(IJ)}= \half(\f_{IJ}+\f_{JI})\ .
\]
The parameter $\LA$ of the compensating supergauge transformation is
given by
\[
\LA= \e^I\lp \theta^J \f_{(IJ)} + \dfrac{1}{4}\theta^2 \eta_I \rp \ ,
\]
and the modified supersymmetry transformations are
\[\ba{l}
\tQ _{I}a=\psi _{I}\ ,\quad 
\tQ _{I}\psi _{J}=-D(a)\phi _{IJ}-\epsilon _{IJ}\alpha\ ,\quad
\tQ _{I}\alpha = \e^{JK}\left[ \phi_{IJ},\psi _{K}\right]
+ D(a)\eta _{I}\ ,\es
\tQ _{I}\phi _{JK}=\half \lp \epsilon_{IJ}\eta _{K}
  +\epsilon _{IK}\eta _{J} \rp\ ,
\quad \tQ_I\eta_J = \e^{KL}\lc \f_{IK},\f_{JL}\rc   \ .
\ea\]
The superalgebra closes on the $\f$ dependent gauge transformations
$\d_{\f_{(IJ)}}$: 
\[
[\tQ_I, \tQ_J] = -2\d_{\f_{(IJ)}}\ .
\]
\end{example}

%%%%%%%%%%%%%%%%%%%%%%%%%%%%%%%%%%%%%%%%%%%%%%%%%%%%%%%%%%
\section{Actions}\label{actions}

%%%%%%%%%%%%%%%%%%%%%%%%%%%%%%%%%%%%%%%%%%%%%%%%%%%%%%%%%%
\subsection{Action for $N_T=1$ in  $D$-dimensions}

\subsubsection{The geometrical sector}
We follow here~\cite{horne,bbrt-pr,blau-thom-96,bbt-ijmp}.
In such theories, the action is purely of gauge fixing type, 
the gauge condition being that of zero Yang-Mills curvature -- or
possibly of selfdual curvature, in four dimensions, as in the 
original Witten's
paper~\cite{witten-donald}. The ``gauge invariance'' which 
has to be fixed is the local shift supersymmetry expressed 
by the nilpotent operator $Q$ (or $\tQ$ in the WZ gauge).
For this we have to introduce a Lagrange multiplier 
field\footnote{Recall that the indices $p$ and $s$ in
$\;^s\vf_p^g$ respectively denote the form degree and the SUSY-number.
The indice $g$ denotes the ghost number associated to the BRST 
invariance defined by \equ{BRST-Transform} -- 
or \equ{brst}, in the WZ gauge.}
$\;^{0}b_{D-2}^0$ and an associated ``antighost'' $\;^{-1}\bb_{D-2}^0$. 
In the case of a selfduality condition in $D=4$ dimensions, both 
$\;^{0}b_{D-2}^0$ and $\;^{-1}\bb_{D-2}^0$ are to be taken as anti-selfdual 
2-forms.
One has still to introduce the Lagrange multiplier $\;^{-2}\bfi_0^0$ 
and its associated ``antighost'' $\;^{-1}\eta_0^0$ in order to fix the 
zero mode of the 1-form field $^1\p_1^0$. ``Antighosts'' and Lagrange 
multipliers transform as 
\eq\ba{ll}
Q \;^{-1}\bb_{D-2}^0 = \;^{0}b_{D-2}^0\ ,
&\quad Q \;^{0}b_{D-2}^0 = 0\ ,\es
Q \;^{-2}\bfi_{0}^0 = \;^{-1}\eta_{0}^0\ ,
&\quad Q \;^{-1}\eta_{0}^0 = 0\ .
\ea\eqn{s-lag-transf-n1}
The best way to write down an invariant action is to use the
superspace formalism, introducing
the two ``Lagrange multiplier superfields''
\[
\;^{-1} B_{D-2}^0 = \;^{-1}\bb_{D-2}^0 + \theta \;^{0}b_{D-2}^0\ ,\quad
\;^{-2}\BFI_{0}^0 = \;^{-2}\bfi_{0}^0 + \theta \;^{-1}\eta_{0}^0\ ,
\]
corresponding to the transformation rules \equ{s-lag-transf-n1}. One must
impose the anti-selduality condition
$P_+\;^{-1} B_{D-2}^0=0$ if one is interested in 
the instanton configurations (see \equ{selfduality} for the 
definition of the (anti-)selfduality projectors).
An action which fixes local shift supersymmetry may be given by the following 
supergauge invariant and supersymmetric expression, written as a 
superspace integral:
\eq\ba{l}
S_{\rm inv} = \tr\dint d^1\te\; 
\lp \;^{-1} B_{D-2}^0 \;F(A) + \;^{-2}\BFI_{0}^0 \;D(A)*\Psi \rp
\es\phantom{S_{\rm inv}}
= \tr\dint
\Lp \;^{0}b_{D-2}^0 \;F(a) + \;^{-1}\eta_{0}^0 \;D(a)*(\p + D(a)\chi) 
 + \;(-1)^{D-1} \;^{-1}\bb_{D-2}^0 \;D(a)\psi
\es\phantom{S_{\rm inv}= \tr\dint\Lp}
+  \;^{-2}\bfi_{0}^0 \Lp - \;D(a)*( D(a)\f + [\p,\chi] )
+ \lc\p\,,*(\p+D(a)\chi)\rc \Rp  \Rp \ ,
\ea\eqn{inv-action-n1}
where * is the Hodge duality symbol. In the second term of the first line, 
we have used 
the supercurvature component $\Psi$ given in \equ{supercampo3}
instead of $\psi$ for the sake of supergauge invariance. In the WZ gauge, 
$\chi=0$, we have
\eq\ba{l}
S_{\rm inv} = 
\tr\dint
\Lp \;^{0}b_{D-2}^0 \;F(a) + \;^{-1}\eta_{0}^0 \;D(a)*\p
 + \;(-1)^{D-1} \;^{-1}\bb_{D-2}^0 \;D(a)\psi
\es\phantom{S_{\rm inv}= \tr\dint\Lp}
-  \;^{-2}\bfi_{0}^0 \;D(a)* D(a)\f
+ \lc\p\,,*\p\rc \Rp \ .
\ea\eqn{inv-action-n1-WZ}

Beyond the zero-mode of the connection superfield $A$ due to super-Yang-Mills 
invariance \equ{BRST-comp}, there still remains 0-modes
for the $(D-2)$-form superfield $\;^{-1} B_{D-2}^0$, due to an invariance 
under local transformations of the form
\eq
\d\;^{-1} B_{D-2}^0 = D(A) \;^{-1}\S_{D-3}^0\ .
\eqn{bf-type-transf}
Before describing our way of fixing these zero-modes, let us briefly
remind of the scheme introduced in~\cite{blau-thom-96}.

%%%%%%%%%%%%%%%%%%%%%%%%%%%%%%%%%%%%%
\subsubsection{The Blau-Thompson gauge fixing}\label{B-T-gauge-fixing}

The fixing of the zero-modes of $\;^{-1} B_{D-2}^0$ by
the authors of~\cite{bbrt-pr,blau-thom-96,bbt-ijmp}
is based on the Batalin-Vilkovisky
procedure~\cite{bv}, adapted to the case where gauge invariance is the 
shift symmetry, with a corresponding system of ghosts for ghosts, antighosts and
Lagrange multipliers. The result 
is rather cumbersome and redundant, but the authors of \cite{blau-thom-96}
succeeded to construct a reduced procedure with a minimum number of
fields. The reduced procedure amounts to introduce a set of superfields,
which we shall denote by 
\eq\ba{l}
 \;^{0}\BP^{0}_{D-3}\ ,\quad  
\;^{-1}\BP^{0}_{D-4}\ ,\quad \;^{0}\BP^{0}_{D-5}\ ,\quad
\cdots\ ,\quad \;^{-k}\BP^{0}_{0}\ ,\quad%\es
%\qquad\qquad\qquad\qquad
\mbox{with}\ k=\frac{1}{2}\lp 1 + (-1)^D\rp   \ ,
\ea\eqn{BT-fields}
and to add to the action \equ{inv-action-n1} the terms
\eq\ba{l}
S_{\rm BT} = \tr\dint d^1\te\; 
\Lp \;^{0}\BP_{D-3}^0 \;D(A)*\;^{-1}B_{D-2}^0 
\;+ \;^{-1}\BP_{D-4}^0 \;D(A)*\;^{0}\BP_{D-3}^0 
\es\phantom{S_{\rm BT}= \tr\dint d^1\te\;}
\;+ \;^{0}\BP_{D-5}^0 \;D(A)*\;^{-1}\BP_{D-4}^0 
\;+ \cdots
+ \;^{-k}\BP_{0}^0 \;D(A)*\;^{k-1}\BP_{1}^0 \Rp \ ,
\ea\eqn{blau-th-action}
which by construction is a $Q$-variation. 
If supplemented by a gauge fixing action for the Yang-Mills supergauge
invariance, the fixing of the zero-modes is complete, propagators are
well defined and the quantum theory may be calculated. However, the
latter is not defined unambiguously. This can be seen, at the
perturbative level, from the possible occurrence of gauge invariant and 
supersymmetric counterterms
different from the terms already present in the action. For instance, in
$D=4$ dimensions, possible such counterterms are
given by superspace integrals of traces of expressions 
such as 
\eq
\;^{0}\BP^0_1 D(A)\;^{-1}B^0_2\ ,\quad 
\;^{0}\BP^0_1 \;^{-2}\BFI^0_0*\Psi\ ,\quad 
\;^{-1}\BP^0_1 \;^{-2}\BFI^0_0*\Phi\ ,\quad 
\;^{-1}B^0_2 \lp\dth\;^{-1}B^0_2+[E,\;^{-1}B^0_2]\rp \ ,\quad \mbox{etc.} 
\eqn{counterterms-to-BT}
This fact may jeopardize the stability of the theory
under radiative corrections.  

Let us remind that there is an alternative way~\cite{horne}, which 
may be used in 
the instanton configuration case, in $D=4$ dimensions. It
consists in adding to the action 
\equ{inv-action-n1-WZ}, instead of the terms \equ{blau-th-action},
a term quadratic in the Lagrange multiplier $\;^{-1}B_{D-2}^0$:
\eq
\half\tr\dint d^1\te \lp \;^{-1}B_{2}^0\;* 
\lp\dth\;^{-1}B^0_2+[E,\;^{-1}B^0_2]\rp  \rp \ , 
\eqn{quad-lagr-mult}
equal to
\eq
\half\tr\dint \lp\;^0b_{2}^0\;*\;^0b_{2}^0
 + \;^{-1}\bb^0_2[\;^{-1}\bb^0_2,\f] \rp \ ,
\eqn{quad-lagr-mult-wz}
in the WZ gauge,
and substituting the now auxiliary field $\;^0b_2^0$ by its equation of
motion
$\;^0b_2^0 = P_- F(a)$, where $P_-$ is the anti-selfduality projector
defined in \equ{selfduality}. This leads to the term
\[
S_{\rm H}=\half\tr\dint \lp P_-F(a) \rp^2\ ,
\]
as pointed out in~\cite{horne}, thus leading to Witten's original
action\footnote{This point is discussed in~\cite{bellisai} together 
with an argument indicating the equivalence of both versions.}. 
This alternative way is analogous to the way leading from a 
gauge fixing of the 
Landau type to one of the Feynman type in usual gauge theories. 
We note that the action $S_{\rm inv}+S_{\rm H}$
represents a complete gauge fixing, too, since the
$BF$-type gauge invariance is explicitly broken.
Moreover, it is stable under the radiative
corrections, to the contrary of the action $S_{\rm inv}+S_{\rm BT}$. 
However, this alternative procedure appears unsuitable for
generalization to higher dimension and higher supersymmetry. 

On the other hand, the
reduced Blau-Thompson procedure may be easily generalized to 
higher dimension and higher $N_T$
shift supersymmetry: this has been 
%explicitly 
done 
in~\cite{blau-thom-96}
for $D=$ 3 and 4, $N_T=$ 1 and 2.
However the same problem of unstability will persist.

%%%%%%%%%%%%%%%%%%%%%%%%%%%%%%%%%%%%%
\subsubsection{The super-$BF$ gauge fixing}

Our proposal is to treat the theory as a supersymmetric theory with
supergauge invariance, and to eliminate the zero-modes of the superfield
$\;^{-1} B_{D-2}^0$ by explicitly using the supergauge invariance
of the type encountered in topological $BF$ theories and fixing it 
accordingly to the Batalin-Vilkovisky (BV) prescription~\cite{bv}, as
in $BF$ theories. Implementing this new gauge invariance within the BRST
algebra, we first introduce
a ghost $\;^{-1}B_{D-3}^1$ as well as
a series of ghosts for ghost $\;^{-1}B_{D-2-g}^g$, $g=2,\dots,D-2$,
and the BRST transformation rules
\eq\ba{l}
\cas \;^{-1} B_{D-2}^0 = -[C,\;^{-1}B_{D-2}^{0}] - D(A) \;^{-1}B_{D-3}^0\ ,\es
\cas \;^{-1}B_{D-2-g}^g = -[C,\;^{-1}B_{D-2-g}^{g}] - D(A) \;^{-1}B_{D-3-g}^{g+1}
\quad (g=1,\dots,D-3)\ ,\es
\cas \;^{-1}B_{0}^{D-2} = -[C,\;^{-1}B_{0}^{D-2}]\ ,
\ea\eqn{brst-bf-n1}
where we have incorporated the super-Yang-Mills transformations with superghost
$C$. We note that, if the space-time dimension $D$ is greater or equal to 4, 
 these transformations hold only on-shell,
namely modulo terms linear in the curvature $F(A)$, 
the latter being an equation of motion as a
consequence of the action \equ{inv-action-n1}. Indeed, $\cas^2=0$ when
applied to all the fields, except
\eq
\cas^2 \;\;^{-1}B_{D-2-g}^g\; =  - \lc F(A), \;^{-1}B_{D-4-g}^{g+2} \rc
\quad (g=1,\dots,D-3\ ;\quad D\ge4)\ .
\eqn{on-shell-nilpot}
The transformations as written in \equ{brst-bf-n1}
hold in the generic case describing the gauge field
configurations of null curvature: $F(a)=0$. If we are interested in the selfdual
configurations  in four-dimensional space-time, $P_-F(a)=0$, 
the Lagrange multiplier superfield $^{-1} B_{2}^0$ has to be chosen as an 
anti-selfdual 2-form:
\eq
P_+ \,^{-1} B_{2}^0=0, \ ,
\eqn{anti-sef-cond}
and the 
BRST transformations \equ{brst-bf-n1} must be redefined accordingly:
\eq\ba{l}
\cas \;^{-1} B_{2}^0 = -[C,\;^{-1}B_{4}^{0}] 
- P_-\lp D(A) \;^{-1}B_{1}^0\rp \ ,\es
\cas \;^{-1}B_{1}^1 = -[C,\;^{-1}B_{1}^{1}] - D(A) \;^{-1}B_{0}^{2}\ ,\es
\cas \;^{-1}B_{0}^{2} = -[C,\;^{-1}B_{0}^{2}]\ .
\ea\eqn{brst-bf-n1-selfdual}
One readily verifies that on-shell nilpotency still holds, $F(A)$ in 
\equ{on-shell-nilpot} being replaced by $P_-F(A)$, which is now
the relevant equation of motion.

The fixing of the gauge invariance \equ{bf-type-transf} 
is completed 
through the addition of antighost and Lagrange multiplier superfields 
$\;^s\BC_p^{g-1}$ and $\;^s\Pi_p^{g}$. The ghosts $B$ and antighosts $BC$
form together a Batalin-Vilkovisky triangle, whose upper summit is the 
superfield $\;^{-1} B_{D-2}^0$ and bottom line is made of 0-forms:
{\small \[
\ba{lllllllll}
&&&&\;^{-1} B_{D-2}^0&&&&\\[5mm]
&&&\;^{0}\BC^{-1}_{D-3}&&\;^{-1}B^1_{D-3}&&&\\[5mm]
&&\;^{-1}\BC^0_{D-4}&&\;^{0}\BC^{-2}_{D-4}&&\;^{-1}B^2_{D-4}&&\\[5mm]
&\;^{0}\BC^{-1}_{D-5}&&\;^{-1}\BC^1_{D-5}&&\;^{0}\BC^{-3}_{D-5}
&&\;^{-1}B^3_{D-5}&\\[5mm]
\cdots&&\cdots&&\cdots&&\cdots&&\cdots
\ea
\] }
The Lagrange multipliers form a smaller triangle corresponding to the 
antighost subtriangle:
{\small \[
\ba{lllllll}
&&&\;^{0}\Pi_{D-3}^0&&&\\[5mm]
&&\;^{-1}\Pi^{1}_{D-4}&&\;^{0}\Pi^{-1}_{D-4}&&\\[5mm]
&\;^{-0}\Pi^{0}_{D-5}&&\;^{-1}\Pi^{2}_{D-5}&&\;^{0}\Pi^{-2}_{D-5}&\\[5mm]
\cdots&&\cdots&&\cdots&&\cdots\\[3mm]
\ea
\] }
The set of BRST transformations given by \equ{BRST-comp} 
for the connection superfields
$A$ and $E$, by \equ{brst-bf-n1} for $\;^{-1} B^0_{D-2}$ and its
ghosts, is
completed by 
\eq
\cas \;^{s}\BC^{g-1}_{p} = \;^{s}\Pi^{g}_{p}\ ,\quad
 \cas \;^{-1}\Pi^{g}_{p}= 0 \ ,
\eqn{BRST-lagr-n1}
for the antighost and Lagrange multipliers, and finally by
\eq
\cas \;^{-2}\BFI_{0}^0 = -[C,\;^{-2}\BFI_{0}^{0}] \ ,
\eqn{BRST-BFI-n1}
the nilpotency property being preserved.
Introducing still the antighost and Lagrange multiplier superfields
$\BC$ and $\Pi$ for fixing super-Yang-Mills gauge invariance, we are ready 
to write down a complete action. Since the
``$BF$ gauge symmetry'' algebra is closed only on-shell, one must use the complete 
Batalin-Vilkovisky setting, including the introduction of the antifields,
demand that the action solves the master equation, thereby obtaining an
action involving terms quadratic in the ghosts. This has been done in
quite generality for the usual $BF$ models~\cite{bbrt-pr,lucchesi-bf}
and will not be repeated here. We shall only indicate the part, 
written as a superspace integral, of the 
action linear in the ghost fields, which may be obtained adding to the 
invariant action \equ{inv-action-n1} a BRST variation:
\eq\ba{l}
S_{\rm (linear\ part\ in\ the\ ghosts)} = S_{\rm inv} \es
 - \cas\;\tr\dint d^1\te\;\Lp \BC\; d*A +
\;^{0}\BC^{-1}_{D-3} \;d*^{-1} B_{D-2}^0 +
\;^{0}\BC^{-2}_{D-4} \;d*^{-1}B_{D-3}^1 \es
\phantom{ - \cas\;\tr\dint d^1\te\;\Lp}
+ \;^{-1}\BC^{0}_{D-4} \;d*^{0}\BC_{D-3}^{-1} + \cdots \Rp\es
 = \tr\dint d^1\te\Lp 
\;^{-1} B_{D-2}^0 \;F(A) + \;^{-2}\BFI_{0}^0 \;D(A)*\Psi
+ \Pi\; d*A - \BC\; d*\cas A \es
\phantom{\tr}  +
\;^{0}\Pi^{0}_{D-3} \;d*\;^{-1} B_{D-2}^0 +
\;^{0}\Pi^{-1}_{D-4} \;d*\;^{-1}B_{D-3}^1 +
 \;^{-1}\Pi^{1}_{D-4} \;d*\;^{0}\BC_{D-3}^{-1} \es
\phantom{\tr} 
- \;^{0}\BC^{-1}_{D-3} \;d* \cas \;^{-1} B_{D-2}^0 
- \;^{0}\BC^{-2}_{D-4} \;d* \cas \;^{-1}B_{D-3}^1 
-  \;^{-1}\BC^{0}_{D-4} \;d* \cas \;^{0}\BC_{D-3}^{-1}
+\cdots \Rp \ .
\ea\eqn{action-n1}
In fact, the dependence in the Lagrange multipliers is exact and
completely fixed if one imposes, as it may be done in usual gauge 
theories~\cite{44}, the Landau type ``gauge conditions''
\eq\ba{l}
\dfud{S}{\Pi} = d*A\ ,\quad 
\dfud{S}{\;^0\Pi^0_{D-3}} = d*\;^{-1} B_{D-2}^0\ ,\es
\dfud{S}{\;^0\Pi^{-1}_{D-4}} = d*\;^{-1} B_{D-3}^1\ ,\quad
\dfud{S}{\;^{-1}\Pi^1_{D-4}} = d*\;^{0} \BC_{D-3}^{-1}\ ,\quad
\cdots\ ,
\ea\eqn{gauge-cond}
which, being linear, are not subject to renormalization.

For the sake of completeness, let us write the expansions of 
the various superfields present in this action:
\eq\ba{l}
A=a+\theta\p\ ,\quad E=\chi+\theta\f\ , \quad
 \;^{-1}B_{D-2-g}^g = \;^{-1}\bb_{D-2-g}^g +\theta \;^{0}b_{D-2-g}^g 
\ ,\es 
\;^{-2}\BFI_{0}^0 = \;^{-2}\bfi_{0}^0 + \theta \;^{-1}\eta_{0}^0 \ ,\quad
\Pi= \pi'+\theta\pi\ ,\quad
\;^{s}\Pi^{g}_{p}=\;^{s}(\pi')^{g}_{p}+\theta \;^{s+1}\pi^{g}_{p} \ ,\es
\;^{s}\BC^{g}_{p}=\;^{s}(\bc')^{g}_{p}+\theta \;^{s+1}\bc^{g}_{p}\ .
\ea\eqn{s-field-exp-n1}

%%%%%%%%%%%%%%%%%%%%%%%%%%%%%%%%%%%%%%%%%%%%%%%%%%%%
\begin{example}\label{ex-D=3} {\bf -- Case $D=3$}

The BRST operator $\cas$ is strictly nilpotent, and the complete action reads
\eq\ba{l}
S  = \tr\dint d^1\te\Lp 
\;^{-1} B_{1}^0 \;F(A) + \;^{-2}\BFI_{0}^0 \;D(A)*\Psi \es
\quad\phantom{\tr} + \Pi\; d*A - \BC\; d*\cas A +
\;^{0}\Pi^{0}_{0} \;d*\;^{-1} B_{1}^0  
- \;^{0}\BC^{-1}_{0} \;d* \cas \;^{-1} B_{1}^0  \Rp\ ,
\ea\eqn{action-n1-d3}
which, in component fields, yields
(see \equ{supercampo3} for the $\theta$-expansion of $\Psi$)
\eq\ba{l}
S
= \tr\dint\Lp 
\;^{0}b_{1}^0 \;F(a) + \;^{-1}\eta_{0}^0 \;D(a)*(\psi+D(a)\chi) 
- \;^{-1}\bb_{1}^0 \;D(a)\p
\es \phantom{S=\tr\dint\Lp}
%%%%%%%%%%%%%%%%%%%%%%%%%%%%%%%%%%%%%
   + \;^{-2}\bfi_{0}^0 \lp\;D(a)*(D(a)\f-[\p,\chi]) 
+ [\p,*(\psi+D(a)\chi)] \rp 
\es \phantom{S=\tr\dint\Lp}
%%%%%%%%%%%%%%%%%%%%%%%%%%%%%%%%%%%%%
 + \pi\; d*a   + \pi'\; d*\p +  \;^{1}\pi^{0}_{0} \;d*\;^{-1}\bb_{1}^0 
 \;- \;^{0}(\pi')^{0}_{0} \;d*\;^{0} b_{1}^0  
\es\phantom{=\tr\dint\Lp}
 %%%%%%%%%%%%%%%%%%%%%%%%%%%%%%%%%%%%%%%%%%%%%%%%%%%%
- \bc\; d*\cas a - \bc'\; d*\cas \p 
 - \;^{1}\bc^{-1}_{0} \;d* \cas \;^{-1}\bb_{1}^0 
 \;+  \;^{0}(\bc')^{-1}_{0} \;d* \cas \;^{0} b_{1}^0   \Rp\ ,
\ea\eqn{action-comp-n1-d3}
In the WZ gauge $\chi=0$, this gives
\eq\ba{l}
S
= \tr\dint\Lp 
\;^{0}b_{1}^0 \;F(a) + \;^{-1}\eta_{0}^0 \;D(a)*\psi 
- \;^{-1}\bb_{1}^0 \;D(a)\p
   + \;^{-2}\bfi_{0}^0 \lp\;D(a)*D(a)\f + [\p,*\p] \rp 
\es \phantom{S=\tr\dint\Lp}
%%%%%%%%%%%%%%%%%%%%%%%%%%%%%%%%%%%%%
 + \pi\; d*a   + \pi'\; d*\p +  \;^{1}\pi^{0}_{0} \;d*\;^{-1}\bb_{1}^0 
 \;- \;^{0}(\pi')^{0}_{0} \;d*\;^{0} b_{1}^0  
\es\phantom{=\tr\dint\Lp}
 %%%%%%%%%%%%%%%%%%%%%%%%%%%%%%%%%%%%%%%%%%%%%%%%%%%%
- \bc\; d*\cas a - \bc'\; d*\cas \p 
 - \;^{1}\bc^{-1}_{0} \;d* \cas \;^{-1}\bb_{1}^0 
 \;+  \;^{0}(\bc')^{-1}_{0} \;d* \cas \;^{0} b_{1}^0   \Rp\ ,
\ea\eqn{action-comp-n1-d3-WZ}
On may observe that the latter action contains the term 
$\pi'd*\p$ which, compared with the term $\;^{-1}\eta_{0}^0 \;D(a)*\psi$, 
shows that the fields $\pi'$ and $\;^{-1}\eta_{0}^0$ are redundant 
and the quadratic part of the action, singular. This
redundancy is an artifact of having written the action in the WZ 
gauge, where $\chi=0$. In the supersymmnetric gauge yielding the action
\equ{action-comp-n1-d3}, 
the field $\;^{-1}\eta_{0}^0$ also couples to $\chi$,
and there is therefore no redundancy. When restricting to the WZ gauge, in order
to get rid of this redundancy, one has to put $\pi'=0$, too.
\end{example}

%%%%%%%%%%%%%%%%%%%%%%%%%%%%%%%%%%%%%%%%%%%%%%%%%%%%
\begin{example}\label{ex-D=4} {\bf -- Case $D=4$}

Let us consider the case of a selfdual curvature, defined by the
anti-selfduality condition \equ{anti-sef-cond} on the $B$-field and the BRST
transformations \equ{brst-bf-n1-selfdual}. The action is 
\eq\ba{l}
S  = \tr\dint d^1\te\Lp 
\;^{-1} B_{2}^0 \;F(A) + \;^{-2}\BFI_{0}^0 \;D(A)*\Psi \es
\quad\phantom{\tr} + \Pi\; d*A  +
\;^{0}\Pi^{0}_{1} \;d*\;^{-1} B_{2}^0 
- \;^{-1}\BC^{0}_{0} \;d*\;^{0} \Pi_{1}^0   \Rp + S_{\rm ghost}\ ,
\ea\eqn{action-n1-d4}
where $S_{\rm ghost}$ is the part of the action depending on the fields
of ghost number $\not=0$, which we shall not write explicitly.
In component fields, in the WZ gauge $\chi=0$ 
and\footnote{See the remark at the end of the preceding example.}
$\pi'=0$, this reads 
\eq\ba{l}
S = \tr\dint\Lp 
\;^{0}b_{2}^0 \;F(a) + \;^{-1}\eta_{0}^0 \;D(a)*\psi 
+ \;^{-1}\bb_{2}^0 \;D(a)\p
   + \;^{-2}\bfi_{0}^0 \lp\;D(a)*D(a)\f + [\p*\p] \rp 
\es \phantom{S=\tr\dint\Lp}
%%%%%%%%%%%%%%%%%%%%%%%%%%%%%%%%%%%%%
 + \pi\; d*a    +  \;^{1}\pi^{0}_{1} \;d*\;^{-1}\bb_{2}^0 
 \;+ \;^{0}(\pi')^{0}_{1} \;d*\;^{0} b_{2}^0  
\es \phantom{S=\tr\dint\Lp}
-  \;^{0}\bc^{0}_{0} \;d*\;^{0}(\pi')_{1}^0 
 \;- \;^{-1}(\bc')^{0}_{0} \;d*\;^{1}\pi_{1}^0   \Rp 
+ S_{\rm ghost}  \ .
\ea\eqn{action-comp-n1-d4}
\end{example}
%%%%%%%%%%%%%%%%%%%%%%%%%%%%%%%%%%%%%%%%%%%%%%%%%%%%
We can see from the actions \equ{action-n1-d3} and 
\equ{action-n1-d4} given in the two examples above, that the 
non-ghost part of the action constructed using the ``super-$BF$ 
gauge fixing'' procedure coincides, in the WZ gauge,
with the action (\ref{inv-action-n1}, \ref{blau-th-action}) 
given by the Blau-Thompson procedure. In $D=4$ dimensions, for instance, 
the Blau-Thompson action is given
by \equ{action-wz-n1d4} and the super-$BF$ like action by 
\equ{action-comp-n1-d4}. They are almost identical, up to changes in 
the notation: 
\[
\;^{1}\la_1 \to \;^{1}\pi^0_1\ ,\quad
\;^{0}\bp_1 \to \;^{0}(\pi')^0_1\ ,\quad
\;^{0}\la_0 \to \;^{0}\bc^0_0\ ,\quad
\;^{-1}\bp_0 \to \;^{-1}(\bc')^0_0\ ,
\] 
and up to the presence of simple derivatives in the latter action
instead of covariant derivatives in the former one. 

In the latter action the
supermultiplets\footnote{Denoted in equations (4.6, 4.7) 
of~\cite{blau-thom-96} by $\{V,\;\bp\}$ and $\{\bar\eta,\;u\}$, respectively.}
 $\{\;^{0}(\pi')^{0}_{1},\;^{1}\pi^{0}_{1}\}$ and 
$\{\;^{-1}(\bc')^{0}_{0},\;^{0}\bc^{0}_{0}\}$
appear naturally as Lagrange multipliers and antighosts  
within the Batalin-Vilkovisky scheme, with couplings fixed uniquely by 
the gauge conditions \equ{gauge-cond}. Hence, due to this and to the gauge
invariance of the $BF$ type defined by \equ{bf-type-transf}, the action 
\equ{action-n1-d4} is uniquely defined -- up to an irrelevant
renormalization of the superfields $\;^{-1} B_{2}^0$ and $\;^{-2}\BFI_{0}^0$,
thus guarantying an unambiguous quantum extension of the theory.
In contrast, 
 $\{\;^{0}(\pi')^{0}_{1},\;^{1}\pi^{0}_{1}\}$ and 
$\{\;^{-1}(\bc')^{0}_{0},\;^{0}\bc^{0}_{0}\}$
appear in the Blau-Thompson approach
as independent supermultiplets introduced together 
with their couplings in an ad hoc way, with the consequence that the action 
(4.7) of~\cite{blau-thom-96} 
is not the most general supersymmetric
and gauge invariant one. Indeed, forgetting the $BF$-type invariance and
the character of Lagrange multiplier and antighost of $\;^{0}\Pi^{0}_{1}$ and 
$\;^{-1}\BC^{0}_{0}$, one would have to consider possible (counter)terms
involving these fields, such as those given by \equ{counterterms-to-BT} 
-- in the notation of subsection \ref{B-T-gauge-fixing} --
which are gauge invariant, supersymmetric and of the same power counting 
dimension 4 as the action.

Of course, these considerations apply as well to the general case of an
arbitrary dimension and also to the models with an arbitrary number of
supersymmetry generators considered in 
next subsection~\ref{extended-susy-actions}.

Let us also repeat that the action as originally given 
by Witten~\cite{witten-donald} 
in the 4-dimensional case, would correspond
to adding to the action \equ{action-comp-n1-d4} the term
\eq
\half\tr\dint d^1\te \lp \;^{-1}B_2^0\; \partial_{\te}\;^{-1}B_2^0 \rp = 
\half\tr\dint \lp\;^0b_2^0\rp^2 \ ,
\eqn{quadr-ct}
and substituting the now auxiliary field $\;^0b_2^0$ by its equation of
motion
$\;^0b_2^0 = -P_- F(a)$, where $P_-$ is the anti-selfduality projector
defined in \equ{selfduality}, thus leading to the term
\[
\half\tr\dint \lp P_-F(a) \rp^2\ .
\]
This would amount to go from a ``gauge fixing'' of the 
Landau type for the local shift symmetry, to one of the Feynman type. 
However, such a term \equ{quadr-ct}
is not allowed in our scheme since it is not invariant under the 
$BF$ type gauge transformation, as we have discussed above.

%%%%%%%%%%%%%%%%%%%%%%%%%%%%%%%%%%%%%%%%%%%%%%%%%%%%%%%
\subsection{Action for Any $N_T$}\label{extended-susy-actions}

The generalization for any number $N_T$ of supersymmetry generators is
straightforward. The $\theta$-expansions of the superfield components 
$A$ and $E_I$ of the superconnection 
$\hA$ \equ{superconexao} and of the superghost $C$ are given in 
(\ref{expansion-A}, \ref{expansions-E-C}). 
Their BRST transformations are given in 
(\ref{BRST-Transform}, \ref{BRST-comp}).
The Lagrange multiplier superfields, 
associated to the zero curvature (or selfduality)
condition and to the fixing of the zero mode of $\p_I$, read 
$\;^{-N_T}B_{D-2}^0$ and $\;^{-N_T-1}(\BFI^I)_0^0 $, respectively.
The supersymmetric and supergauge invariant action is given by
\eq
S_{\rm inv} = \tr\dint\dnth
\lp \;^{-N_T} B_{D-2}^0 \;F(A)
 + \;^{-N_T-1}(\BFI^I)_{0}^0 \;D(A)*\Psi_I \rp \ ,
\eqn{inv-action-n}
whith the supercurvature components $F(A)$ and $\Psi_I$ defined by 
\equ{superComp}.
We shall not spell out this expression, nor the following
ones, in components. 
The ghosts and ghosts for ghost of $\;^{-N_T} B_{D-2}^0$ are shown 
together with their antighosts in the BV triangle
{\small \[
\ba{lllllll}
&&&\;^{-N_T} B_{D-2}^0&&&\\[5mm]
&&\;^{0}\BC^{-1}_{D-3}&&\;^{-N_T}B^1_{D-3}&&\\[5mm]
&\;^{-N_T}\BC^0_{D-4}&&\;^{0}\BC^{-2}_{D-4}&&\;^{-N_T}B^2_{D-4}&\\[5mm]
\cdots&&\cdots&&\cdots&&\cdots
\ea
\] }
and the corresponding Lagrange multipliers in the triangle
{\small \[
\ba{lllll}
&&\;^{0}\Pi_{D-3}^0&&\\[5mm]
&\;^{-N_T}\Pi^{1}_{D-4}&&\;^{0}\Pi^{-1}_{D-4}&\\[5mm]
\cdots&&\cdots&&\cdots
\ea
\] }
The BRST transformations (\ref{BRST-lagr-n1}, \ref{BRST-BFI-n1}) hold, 
and the total action reads, as much 
as its linear part in the ghost fields is concerned:
\eq\ba{l}
S_{\rm (linear\ part\ in\ the\ ghosts)} 
 = \tr\dint\dnth\Lp 
\;^{-N_T} B_{D-2}^0 \;F(A) + \;^{-N_T-1}(\BFI^I)_{0}^0 \;D(A)*\Psi_I 
+ \Pi\; d*A \es
\phantom{\tr} - \BC\; d*\cas A +
\;^{0}\Pi^{0}_{D-3} \;d*\;^{-N_T} B_{D-2}^0 +
\;^{0}\Pi^{-1}_{D-4} \;d*\;^{-N_T}B_{D-3}^1 +
 \;^{-N_T}\Pi^{1}_{D-4} \;d*\;^{0}\BC_{D-3}^{-1} \es
\phantom{\tr} 
- \;^{0}\BC^{-1}_{D-3} \;d* \cas \;^{-N_T} B_{D-2}^0 
- \;^{0}\BC^{-2}_{D-4} \;d* \cas \;^{-N_T}B_{D-3}^1 
-  \;^{-N_T}\BC^{0}_{D-4} \;d* \cas \;^{0}\BC_{D-3}^{-1}
+\cdots \Rp \ ,
\ea\eqn{action-n}
with $F(A)$ and $\Psi_I$ given by \equ{supercampo3-N=2}.
The couplings of the Lagrange multipliers are still defined by the gauge
conditions \equ{gauge-cond}, with the obvious SUSY-number substitution $-1$
$\to$ $-N_T$ in due place.
\newpage

%%%%%%%%%%%%%%%%%%%%%%%%%%%%%%%%%%%%%%%%%%%%%%%%%%
\begin{example}  {\bf -- Case $N_T=2$, $D=3$}

The complete action is
\eq\ba{l}
S = \tr\dint d^2\te\Lp 
\;^{-2} B_{1}^0 \;F(A) + \;^{-3}(\BFI^I)_{0}^0 \;D(A)*\Psi_I
\es \phantom{S = \tr\dint d^2\te\Lp } 
+ \Pi\; d*A - \BC\; d*\cas A
 + \;^{0}\Pi^{0}_{0} \;d*\;^{-2} B_{1}^0  
- \;^{0}\BC^{-1}_{0} \;d* \cas \;^{-2} B_{1}^0 
 \Rp \ .
\ea\eqn{action-n2-d3}
We can write this action in components, in the WZ gauge
$\chi_I=0$, $\f_{IJ}-\f_{JI}=0$, using
the $\te$-expansions defined in \equ{exp-N=2} and by
\eq\ba{ll}
^{-1}B_{D-2-g}^g = \bb(x) +\te^I b_I(x) +\half\te^2 b(x) \ ,\qquad
&^{-2}(\BFI^I)_{0}^0 = \bfi^I + \theta^J\bfi^I_J +\half\te^2 \bfi^I_F\ ,\es
\Pi= \pi + \theta^I\pi_I +\half\te^2 \pi_F\ ,\qquad
&^0\Pi^0_0 = \pi' + \theta^I\pi'_I +\half\te^2 \pi'_F\ ,
\ea\eqn{s-field-exp-n2-d3}
The result is, restricted to the quadratic terms: 
\eq\ba{l}
S_{\rm quadr} = - \tr\dint\Lp bf(a) -\e^{IJ}b_Id\p_J + \bb d\a
  +\bfi_F^Id*\p_I + \e^{JK} \bfi^I_Jd*(\e_{IJ}\a+d\f_{IJ}) + \bfi^Id*d\eta_I
\es\phantom{S_{\rm quadr} = - \tr\dint\Lp}
+\pi_F d*a + \e^{IJ}\pi_I d*\p_J + \pi d*\a
  +\pi'_F d*\bb + \e^{IJ}\pi'_I d*b_J + \pi' d*b \Rp\ .
\ea\eqn{action-n2-d3-comp}
As in the $N_T=1$ case, one has redundancy in some of the fields, which must be
eliminated by putting $\pi=\pi_I=0$.

One can see that this action -- like in the examples 
\ref{ex-D=3} and \ref{ex-D=4} --  also
corresponds to an action written by Blau and Thompson 
(eq. (4.5) of~\cite{blau-thom-96}).  
\end{example}

%%%%%%%%%%%%%%%%%%%%%%%%%%%%%%%%%%%%%%%%%%
\section{Examples of Observables}\label{observables}

It has been shown in~\cite{boldo} that all the observables for $N_T=1$,
defined as BRST cohomology classes of supersymmetric field polynomials,
are given from the Chern classes associated to the superconnection $\hA$
\equ{superconexao}, and that the result is equivalent to the result of 
Witten given in subsection \ref{observ-de-Witten}. We shall give here
the generalization   for any value
of $N_T$, however without proving that
this still gives the complete set of observables~\cite{boldo-prepa}.
The observables are completely determined from the general solution of the 
superdescent equations
\eq\ba{l}
\cas \hO_D + \hd\hO_{D-1}^1=0\ ,\quad
\cas \hO_{D-1}^1 + \hd\hO_{D-2}^2=0\ ,\quad\cdots \ ,\quad
\cas \hO_{0}^{D}=0 \ .
\ea\eqn{s-descent}
where  $\hO_D(x,\theta)$ are superforms of ghost number 0 and 
superform degree $D$ which are nontrivial elements of 
the cohomology $H(\cas|\hd)$
of $\cas$ modulo $\hd$ in the space of the superforms, 
\[\ba{l}
\cas\hO = 0\quad(\mbox{modulo\ }\hd)\ ,\quad
\mbox{but}\quad
\hO\not= \cas{\hat\F}\quad(\mbox{modulo\ }\hd)\ .
\ea\]
Expanding  $Q^{N_T}\hO_D = (\dth)^{N_T}\hO_D$ according to the space-time 
form degree $p$:
\eq
Q^{N_T}\hO_D = \sum_{p=0}^{D}   
 w_{p;\,I_1...I_{D-p}} d\theta^{I_1}\cdots d\theta^{I_{D-p}}\ ,
\eqn{gen-observ}
one identifies the space-time forms $w_p$ as the desired solutions. Indeed, 
\[
\cas w_p(x) = 0 \ (\mbox{modulo\ }d)\quad (n\ge1)\ ,
\quad Qw_0 (x) = 0 \ ,
\]
which follows from applying the operator $Q^{N_T}$ to the first of the
superdescent equations \equ{s-descent}, and using the identities
$Q^{N_T}\,\hd=Q^{N_T}\,d = (-1)^{N_T}\,d\,Q^{N_T}$,
which are direct consequences of the definitions.

The general result for \equ{s-descent} is~\cite{boldo}
\eq\ba{l}
\hO_{D} = \theta^{\rm CS}_{r_1}(\hA) 
f_{r_2}(\hF)\cdots f_{r_L}(\hF)\ , \quad
\mbox{with}\quad  D= 2 \, \dsum{i=1}{L} \, m_{r_i} -1 
\ , \quad L \geq 1 \ ,
\ea\eqn{3.xx}
where $f_r(\hF)$ is the supercurvature invariant of degree $m_r$ in 
$\hF$ corresponding to the gauge group Casimir operator of degree
$m_r$, and $\theta^{\rm CS}_{r}(\hA)$ is the associated super-Chern-Simons
form: 
\eq
\hd \theta^{\rm CS}_{r}(\hA) = f_r(\hF)\ .
\eqn{s-CS}
We note that the superform degree of the solution \equ{3.xx} is odd.

%%%%%%%%%%%%%%%%%%%%%%%%%%%%
\begin{example} {\bf -- Maximum degree  $D=3$}

The superdescent equations
read as
\[\ba{l}
\cas \hO_3 + \hd\hO_2^1=0\ ,\quad
\cas \hO_2^1 + \hd\hO_1^2=0\ ,\quad
\cas \hO_1^2 + \hd\hO_0^3=0\ ,\quad
\cas \hO_0^3=0\ .
\ea\]
The unique nontrivial solution is 
\[\ba{l}
\hO_3 =  \tr ( \hA \hd\hA + \frac{2}{3} \hA^3 ) \ ,\quad
\hO_2^1 =  \tr (\hA \hd C ) \ ,\quad
\hO_1^2 =  \tr ( C \hd C ) \ ,\quad
\hO_0^3 =  -\frac{1}{3} \; \tr C^3 \ .
\ea\]
Note that $\hO_3$ is the Chern-Simons superform associated 
to the quadratic Casimir operator of the gauge group.
%\vspace{2mm}
%
Following \equ{s-descent} we get, for $N_T=1$

\eq\ba{ll}
w_0 = \tr(\f^2 + 2\f\chi^2 )\ ,\quad
&w_1 =  2\tr ( \psi\f + \p\chi^2 + 
\f D(a)\chi ) \ ,\es
w_2 = \tr(\p^2 + 2\f F(a) + 2\p D(a)a \chi)\ ,\quad
&w_3 = 2\tr( \p F(a) )  \ .
\ea\eqn{list-fi2}
 The observables are the integrals of these forms 
 (and of $\tr {F(a)} ^2$) 
 on closed submanifolds of appropriate dimension. 

In the Wess-Zumino gauge $\chi=0$:
\[\ba{l}
w_0     =   \tr ( \f^2) \ ,\quad
w_1      =  2\tr ( \p\f  ) \ ,\quad
w_2      = 
\tr ( 2\f F(a) + \p^2) \ ,\quad
w_3      =  2\tr (\p F(a) ) 
\ea\]
which corresponds to Witten's result up to total derivatives.
\end{example}
%%%%%%%%%%%%%%%%%%%%%%%%%%%%%%%%%%%%%%%%%%%%%%%%
For  $N_T=2$ we obtain (in the WZ gauge $\chi_I=0$, $\f_{[IJ]}=0$): 
\[\ba{l}
w_0 = 2\tr \eta_{(I} \f_{JK)}  \ , \es
w_1 =  \tr ( 2 \alpha \f_{(IJ)}  +
               2 \psi_{(I} \eta_{J)}  
             +  \epsilon^{KL} \f_{(I|K}D(a)\f_{L|J)}) \  , \es
w_2 = 2 \tr(\alpha \psi_I + F(a) \eta_I  
         +\epsilon^{JK}\f_{IJ}D(a)\psi_K)\ , \es
w_3 = \tr( 2 \alpha F(a) + \epsilon^{IJ}\psi_I D(a)\psi_J)  .
\ea\]

%%%%%%%%%%%%%%%%%%%%%%%%%%%%%%%%%%%%%%%%%%%%%%%%%%%%%%%%%%%%%%%%
\section{Absence of Radiative Corrections}\label{uv-finiteness}

The Feynman rules in the general case are deduced 
from the action \equ{action-n}.
It is useful to work directly in superspace. The nonzero superpropagadores are
\eq\ba{l}
\Vev{A(1),\;^{-N_T}B^{0}_{D-2}(2)}\ ,\quad
\Vev{A(1),\Pi(2)}\ , \quad\  \Vev{\BC(1), C(2)}\ ,\es
\Vev{E_I(1),\;^{-N_T-1}(\BFI^I)^{0}_{0}(2)}\ ,\quad
\Vev{E_I(1), \Pi(2)}\ ,\es
\Vev{ \;^{-N_T}B^{g}_{D-g-2}(1),\;^{0}\Pi^{-g}_{D-g-3}(2) }\ ,\quad
\Vev{ \;^{-N_T}B^{g}_{D-g-2}(1),\;^{0}\BC^{-g}_{D-g-2}(2) }\quad (g\ge0)\ ,\es
\Vev{ \;^{0}\BC^{-g}_{D-g-2}(1),\;^{-N_T}\Pi^{g}_{D-g-3}(2) }
\quad (g\ge1)\ ,\es
\Vev{ \;^{-N_T}\BC^{g}_{D-g-4}(1),\;^{0}\Pi^{-g}_{D-g-5}(2) }\ ,\quad
\Vev{ \;^{-N_T}\BC^{g}_{D-g-4}(1),\;^{0}\Pi^{-g}_{D-g-3}(2) }\quad(g\ge0)\ ,\es
\Vev{ \;^{0}\BC^{-g}_{D-g-4}(1),\;^{-N_T}\Pi^{g}_{D-g-5}(2) }\ ,\quad 
\Vev{ \;^{0}\BC^{-g}_{D-g-4}(1),\;^{-N_T}\Pi^{g}_{D-g-3}(2) }\quad(g\ge0)\ ,\es
\mbox{etc.}\ ,
\ea\eqn{prop-list}
where we are using the notation $\vf(n)$ for $\vf(x_n,\te_n)$. 
With one irrelevant exception  shown hereafter, 
all these propagators have as factor a
$\te$-space $\d$-function $(\te_1-\te_2)^{N_T}$. For instance, the first
one reads
\eq
\vev{A_\m(1),\;^{-N_T}B^{0}_{\n_1\cdots \n_{D-2}}(2)}
\sim \D^{-1}\; \e_{\m\n_1\cdots \n_{D-2}\rho}\pa^\rho \d(1,2)\ ,
\eqn{A-B}
up to some numerical factor, where 
$\D^{-1}$ is the inverse of the Laplace operator $\D=*d*d + d*d*$, and
\[
\d(1,2)= \d^D(x_1-x_2) \dfrac{(-1)^{N_T+1}}{N_T!}(\te_1-\te_2)^{N_T}
\]
is the $(D,N_T)$-superspace Dirac distribution. The exception is the
propagator
\eq
\vev{E_I(1),\;\Pi(2)}
\sim \D^{-1}\; \pad{}{\te^I_1} \d(1,2)\ ,
\eqn{E-Pi}
which is of degree $N_T-1$ in $\te_1-\te_2$. However the latter does not contribute 
to any 1-particule irreducible (1PI) graph since the Lagrange multiplier 
superfield $\Pi$
has no interaction in virtue of the gauge conditions \equ{gauge-cond}. 

Now, repeating a well known argument of superspace 
diagrammatic~\cite{1001,pig-sib}, we observe that,
since all contributing propagators have a factor 
$(\te_m-\te_n)^{N_T}$, the integrant of a nontrivial 
1PI graph with $N$ vertices will 
be homogeneous of degree $N\times N_T$ in the differences $\te^I_m-\te^I_n$. 
On the other hand, having $N\times N_T$
independent Grassmann coordinates, we can only form $(N-1)\times N_T$
independent differences. Hence, due to the anticommutativity of the
$\te$'s, the integrant will vanish. 
We thus conclude to the complete absence of 
radiative corrections.

%%%%%%%%%%%%%%%%%%%%%%%%%%%%%%%%%%%%%%%%%%%%%%%%%%%%%%%%%%%%%%%%
\section{Conclusion}

We have developped a general scheme, based on superspace formalism,
which allows for a systematic construction of topological Yang-Mills
theories for arbitrary numbers of shift supersymmetry generators and
space-time dimensions. The main advantage of this scheme, beyond its
systematic character, is that it leads
to an unambiguous determination of the respective actions, thanks to the
introduction of a $BF$ theory type supergauge invariance, which has 
been fixed accordingly to the Batalin-Vilkovisky prescriptions.
Moreover, the ultraviolet finiteness --  in fact 
the absence of radiative
corrections -- follows, in the supersymmetric gauge fixing we have
chosen, directly from the superspace Feynman rules.

%%%%%%%%%%%%%%%%%%%%%%%%%%%%%%%%%%%%%%%%%%%%%%%%%
{\bf Acknowledgments.} We thank  Jos\'e Luis Boldo, Fran\c cois Gieres,
Jos\'e Helay\"el Neto, Matthieu Lefran\c cois and
Jos\'e Alexandre Nogueira
for many useful discussions.
%%%%%%%%%%%%%%%%%%%%%%%%%%%%%%%%%%%%%%%%%%%%%%%%%

%%%%%%%%%%%%%%%%%%%%%%%%%%%%%%%%%%%%%%%%%%%%%%%%%
%\newpage

\appendix

%\section*{Appendices}
\setcounter{section}{0}
%%%%%%%%%%%%%%%%%%%%%%%%%%%%%%%%%%%%%%%%%%%%%%%%%

%%%%%%%%%%%%%%%%%%%%%%%%%%%%%%%%%%%%%%%%%%%%%%%%%

\section*{Appendices. Notations and Conventions}\label{notations}
%%%%%%%%%%%%%%%%%%%%%%%%%%%%%%%%%%%%%%%%%%
\section{Differencial calculus}\label{a-1}

Here, ``space-time'' is an arbitray $D$-dimensional smooth manifold, 
equipped with a Riemannian background metric $(g_{\m\n})$, of determinant
$g>0$.
Space-time objects are differential forms such as $a=a_\m dx^\m$, etc.
We shall call an object even or bosonic (respectively, odd or fermionic) if it
obeys to commutation (respectively, anticommutation) relations.

The bracket $[\cdot,\cdot]$ in general denotes the graded bracket
\eq\ba{l}
[X,Y] = XY+YX\quad\mbox{if both $X$ and $Y$ are odd,}\es
[X,Y] =XY-YX\quad\mbox{otherwise.}
\ea\eqn{graded-bracket}
The fields (forms, superfields, etc.) appearing in this paper are all
taken in the
Lie algebra of the gauge group $G$, which we assume to be compact.
A field  $\vf$ is then a matrix $\vf^a\tau_a$, where
the generators $\tau_a$ obey
the Lie algebra commutation relations and trace property
\eq
[\tau_a,\tau_b] = f_{ab}{}^c \tau_c\ ,\quad \tr \tau_a\tau_b = 2\d_{ab}\ .
\eqn{Lie-alg}
%%%%%%%%%%%%%%%%%%%%%%%%%%%%%%%%%%%%%%%%%%%%%%%%%%
 The {\it Hodge dual} of a $p$-form $\om$ is 
  the $(D-p)$-form $*\om$ defined by
\cite{bertlmann} 
\eq\ba{ll}
&* \om  = \dfrac{1}{(D-p)!}\,\tilde\om{}_{\m_1...\m_{D-p}}
 dx^{\m_1}... dx^{\m_{d-p}} \es
{\rm where} 
\quad &
\tilde\om{}_{\m_1 ... \m_{D-p}} = \dfrac{1}{p!}\,\dfrac{1}{\sqrt{g}}
\, \e_{\m_1 ... \m_D} \om^{\m_{D-p+1}...\m_D}
\ .
\ea\eqn{hodge}
  Here and elsewhere in the text, the wedge product symbol 
has been  omitted. Moreover, 
the background metric $(g_{\m\n})$, as well as the
totally antisymmetric tensor of Levi-Civita:
\eq
\e_{\m_1 ... \m_D} 
= g_{\m_1\n_1} \cdots g_{\m_D\n_D}\e^{\n_1 ... \n_D}\ ,\quad 
\e^{1... D}=1 \ ,\quad \e_{1... D} =  g \ .
\eqn{lev-citta}
The following formulas are quite useful~\cite{bertlmann}:  
\eq
** \om_p = (-1)^{p(D-p)} \, \om_p\ 
\ ,\quad  \omega_p * \phi_p =  \phi_p * \omega_p  \ .
\eqn{dual-comm}
Since the Hodge star operator maps a form of degree $p$ to
a form of total degree $D-p$, it represents an even operator
  if the space-time dimension $D$ is even and an odd operator
otherwise.  
For $D=4$, a {\it selfdual} or {\it anti-selfdual} 2-form
$\om_2$ is defined by the condition $*\om_2=\pm\om_2$. Projectors on 
selfdual or anti-selfdual 2-forms are given by
\eq
P_\pm=\frac{1}{2}(1\pm *)\ .
\eqn{selfduality}

%%%%%%%%%%%%%%%%%%%%%%%%%%%%%%%%%%%%%%%%%%%%%
\section{$N_T$- supersymmetry and superspace}\label{a-2}

$(D,N_T)$-superspace bosonic coordinates are denoted by $x^\m$, 
$\m=0,\dots,D-1$, the fermionic (Grassmann, or anticommuting) 
coordinates being denoted by
$\te^I$, $I-1,\dots,N_T$.
The $N_T$ supersymmetry generators $Q_I$ are represented on superfields
$F(x,\te)$ by
\[
Q_I F = \pa_I F \equiv \dpad{}{\te^I}F\ ,
\]
where, by definition, $\pa_K\te^J=\d_K^J$. Further conventions
 and properties about the
$\te$-coordinates are the following:
\[\ba{l}
\te^{N_T} = \e_{I_1\cdots I_{N_T}}\te^{I_1}\cdots\te^{I_{N_T}}
= N_T!\; \te^1\cdots\te^{N_T}  \ ,\es
(\dth)^{N_T} =\e^{I_1\cdots I_{N_T}}\pa_{I_1}\cdots\pa_{I_{N_T}}
= N_T!\; \pa_1\cdots\pa_{N_T} \ ,\es
(\dth)^{N_T}\te^{N_T} = - (N_T!)^2\ ,

\ea\]
where $\e^{I_1\cdots I_{N_T}}$ 
is the completely antisymmetric tensor of rank $N_T$, with 
the conventions
\[
\e^{1\cdots N_T }=1\ ,\quad 
\e_{I_1\cdots I_{N_T}} = (-1)^{N_T+1} \; \e^{I_1\cdots I_{N_T}}\ .
\]
One may define the conserved supersymmetry number -- {\it SUSY number} -- 
attributing the value 1 to the generators $Q_I$, hence $-1$ to the
$\te$-cordinates. The SUSY number of each field 
component is then deduced from the
SUSY number given to each superfield.

Superspace integration of a superfield form $\OM_p(x,\te)$ is defined
by integrals
\[
\dint\dnth \OM_p(x,\te) = \dint_{\!\!\!\!\!\!M_p}\;\dint\dnth \OM_p(x,\te)\ ,
\] 
where the $x$-space integral is made on some $p$-dimensional 
(sub)manifold $M_p$, and the $\te$-space integral is the Berezin
integral defined by
\[
\dint\dnth\cdots = - \dfrac{1}{(N_T!)^2}(\dth)^{N_T}\cdots \ ,\quad
\mbox{such that}\quad \dint\dnth \te^{N_T}=1\ .
\]

In the specal case of $N_T=2$, the antisymmetric tensors $\e^{IJ}$ and 
$\e^{IJ}$ may be used for raising and lowering the indices:
\[
\te_I=\e_{IJ}\te^J\ ,\quad \te^I=\e^{IJ}\te_J\ ,
\quad \e_{IJ} = -\e^{IJ}\ ,\quad \e^{12}=1\ ,
\quad \e^{IJ}\e_{JK}=\d^I_K\ ,
\]
and one has the useful formulas 
\[
\te^2 = \te^I\te_I=-\te_I\te^I ,\quad
\te^I\te^J = -\half \e^{IJ}\te^2 \ ,\quad \te_I\te_J = \half \e_{IJ}\te^2\ . 
\]
$N_T=1$ and $N_T=2$ superfields have the conventional expansions
\[\ba{l}
\F(x,\te) = \f(x)+\te \f'(x)\quad (N_T=1)\ ,\es
\F(x,\te) = \f(x)+\te^I \f_I(x) + \half\te^2 \f_F \quad (N_T=2)\ .
\ea\]

\newpage

%%%%%%%%%%%%%%%%%%%%%%%%%%%%%%%%%%%%%%%%%%%%%%%%%%%%%

\end{document}